\RequirePackage{fix-cm}
\documentclass[smallextended]{svjour3}       
\smartqed  
\usepackage{graphicx}
\usepackage{natbib}
\usepackage{algorithm}
\usepackage{multirow}
\usepackage{subfigure}
\usepackage{booktabs}
\usepackage{xcolor}
\usepackage{algpseudocode}
\RequirePackage{amsmath,amssymb}

\makeatletter
\renewcommand{\ALG@name}{Pseudo-algorithm}
\makeatother

\DeclareMathOperator*{\argmax}{arg\,max}
\begin{document}

\title{Better than the best? Answers via model ensemble in density-based clustering
}
\subtitle{}

\titlerunning{Ensemble density-based clustering}        

\author{Alessandro Casa \and 
        Luca Scrucca \and 
        Giovanna Menardi 
}


\institute{Alessandro Casa $\cdot$ Giovanna Menardi \at
              Department of Statistical Sciences, University of Padova  \\
              Via C. Battisti, 241, 35121 Padova, Italy\\
          \email{casa@stat.unipd.it; menardi@stat.unipd.it}          
           \and
           Luca Scrucca \at 
           Department of Economics, University of Perugia \\
           Via A. Pascoli, 20, 06123 Perugia, Italy \\
           \email{luca.scrucca@unipg.it}
}

\date{Received: date / Accepted: date}

\maketitle

\begin{abstract}
With the recent growth in data availability and complexity, and the associated outburst of elaborate modelling approaches, model selection tools have become a lifeline, providing objective criteria to deal with this increasingly challenging landscape. In fact, basing predictions and inference on a single model may be limiting if not harmful; ensemble approaches, which combine different models, have been proposed to overcome the selection step, and proven fruitful especially in the supervised learning framework. Conversely, these approaches have been scantily explored in the unsupervised setting. In this work we focus on the model-based clustering formulation, where a plethora of mixture models, with different number of components and parametrizations, is typically estimated. We propose an ensemble clustering approach that circumvents the single best model paradigm, while improving stability and robustness of the partitions. A new density estimator, being a convex linear combination of the density estimates in the ensemble, is introduced and exploited for group assignment. As opposed to the standard case, where clusters are typically  associated to the components of the selected mixture model, we define partitions by borrowing the modal, or nonparametric, formulation of the clustering problem, where groups are linked with high-density regions. Staying in the density-based realm we thus show how blending together parametric and nonparametric approaches may be beneficial from a clustering perspective.

\end{abstract}

\section{Introduction} \label{sec:intro}

In virtually any scientific domain we are witnessing an explosion in the availability of the data, coupled with a tremendous growth in their complexity. As a straightforward consequence, the number of choices that we have to make is increasing as well as the number of sophisticated modelling strategies proposed to deal with such newly introduced challenges. These choices are practically involved in any phase of the modelling process, spanning a wide landscape of possible options: from choosing a class of models or an appropriate approach to analyze a set of data, to more specific decisions as the selection of subsets of relevant variables or suitable parametrizations. Therefore, nowadays model selection steps, helping to formally extricate ourselves from the labyrinth of all these possible alternatives, are ubiquitous in any data analysis routine. Some commonly considered ways forward hence consist in estimating a set of different models and then selecting the best one according to some criterion \citep{claeskens2008model}, or resorting to penalization schemes aimed at balancing fit and complexity \citep[see][for an introduction]{tibshirani2015statistical}.

Nevertheless basing predictions and inference on a single model could turn out to be sub-optimal. Hence, in order to propose viable alternatives to this paradigm, model averaging and ensemble techniques have been thoroughly studied in literature. Even if these two approaches focus on different phases of the modelling process, respectively estimation and prediction, they share the same founding rationale as they aim to improve performances of the base models by combining their strengths and simultaneously relieving their limits. For this reason the two expressions will be used interchangeably in the rest of the paper. 
With inferential goals in mind, model averaging approaches have been proposed, intended to estimate quantities by computing weighted averages of different estimates. Such strategies may lead to improvements in the estimation process by accounting for model uncertainty. Similarly, from a predictive point of view, ensemble techniques have shown remarkable performances in a lot of different applications by building predictions as combinations of the ones given by a set of different models. Well established methods as  \emph{bagging}, \emph{stacking}, \emph{boosting} or the \emph{random forests} \citep[see][for a review]{friedman2001elements} have become the state of the art in the supervised learning framework.

While extensively studied in the classification context, ensemble techniques have been scarcely pursued in the clustering one. A possible explanation can be found in the unsupervised nature of the problem itself;
the absence of a response variable introduces relevant issues in evaluating the quality of a model and of the corresponding partition. As a consequence, weighting models in order to combine them turns out to be an awkward problem. Nonetheless mixing different partitions in a final one could in principle allows combining clustering techniques based on different focuses to give a multiresolution view of the data and possibly improves the stability and the robustness of the solutions. In this direction \cite{fern2003random} exploit the concept of \emph{similarity matrix} in order to aggregate partitions obtained on multiple random projections, and a similar approach is followed by \cite{kuncheva2004using} to study the concept of diversity among partitions. \cite{monti2003consensus} consider again a similarity matrix in order to evaluate the robustness of a discovered cluster under random resampling. In turn, the work by \cite{strehl2002cluster} introduces three different solutions to the ensemble problem in the unsupervised setting by exploiting hypergraph representations of the partitions.

In this work we focus mainly on the parametric, or model-based, approach to cluster analysis where a one-to-one correspondence between clusters and components of an appropriate mixture model is drawn. Here the usual working routine is based on the  \emph{single best model paradigm}, i.e. a set of models is fitted and only the best one is chosen and considered to obtain a partition. Our aim is to go beyond this paradigm by introducing a model averaging methodology to give partitions resulting from an ensemble of models, thus possibly achieving a greater accuracy and robustness. Averaging is pursued directly on the estimated mixture densities in order to build a new and more accurate estimate. Clustering is then addressed, building on the resulting estimate, within a density-based formulation, yet with a shift to a modal, or nonparametric approach, where clusters are associated to the domains of attraction of the density modes. We turn out with a partition where cluster shapes are not constrained by some distributional assumption, as in the model-based framework, but having arbitrary, possibly non-convex and skewed shapes. Therefore, by combining the strengths of the parametric and nonparametric frameworks, our proposal results in a hybrid method which enjoys the advantages of both. 
 
The rest of the paper is structured as follows. In Section \ref{sec:modelaveraging} we outline the proposed methodology and describe in details the estimation procedure. In Section \ref{sec:chscrucca_discussion} we discuss some specific aspects of our proposal and highlight connections with other models. Lastly in Section \ref{sec:chscrucca_results} we show the performances of our method on both simulated and real datasets, and compare it with some competitors while Section \ref{sec:chscrucca_conclusions} presents some concluding remarks. 

\section{Model averaging in model-based clustering}
\label{sec:modelaveraging}
\subsection{Framework and model specification}
\label{sec:chscrucca_frameworkmodelspec}

The goal of partitioning a set of data into some groups, diffusely known as clustering, has been pursued by proposing a lot of techniques with different rationales behind. While most of them are based on a vague notion of clusters, associated to some measure of similarity, an attempt to obtain a precise formalization of the problem is given by the so called \emph{density-based} approach. Here the concept of cluster finds a formal definition by linking it to some specific features of the density $f : \mathbb{R}^d \rightarrow \mathbb{R}$ assumed to underlie the data $\mathcal{X}=\{x_1,\dots,x_n\}$ and consequently inducing a partition of the whole sample space. Furthermore this assumption allows framing the clustering problem in a standard inferential context where, having a ``ground truth" to aim at, several tools can be used in order to evaluate and compare alternative clustering configurations. 

The idea behind density-based clustering has been developed by taking two distinct paths. In the modal, or nonparametric, clustering formulation, clusters are defined as the ``domains of attraction'' of the modes of the density $f$ \citep{stuetzle2003estimating} usually estimated by means of some nonparametric density estimator \citep[see, for a recent account,][]{chacon2018multivariate}. The operational identification of the modal regions can be addressed following different routes \citep[for a comprehensive review readers can refer to][]{menardi2016review} where the most common one consists in finding explicitly the local maxima of the density by exploiting numerical optimization methods. Most of these methods can be seen as refinements or slight modifications of the mean-shift algorithm \citep{fukunaga1975estimation} that, starting from a generic point, shifts it recursively along the steepest ascend path of the gradient of the density estimate until converging to a mode; the final partition of the data is then obtained by grouping together those observations ascending to the same mode. 

On the other hand the model-based, or parametric, approach \citep{banfield1993model,fraley2002model} represents the other, more widespread, formulation of density-based clustering. In this framework $f$ is assumed to be adequately described by means of finite mixture models. Therefore the density of a generic observation $x_i \in \mathbb{R}^d$ is written as
\begin{equation*}
f(x_i | \Psi) = \sum_{k=1}^K \pi_k \varphi_k(x_i | \theta_k) \; , 
\end{equation*}
where $K$ is the number of mixture components, $\varphi_k(\cdot)$ the $k$th component density, while  $\Psi=(\pi_1,\dots,\pi_{K-1},\theta_1,\dots,\theta_K)$ is the vector of parameters where $\pi_k$ are the mixing proportions with $\pi_k >0, \; \forall k=1,\dots,K$ and $\sum_{k=1}^K \pi_k = 1$. 
When Gaussian densities are employed as mixture components, we may write $\varphi_k(\cdot) = \phi_k(\cdot)$ and $\theta_k = \{\mu_k, \Sigma_k\}$. The concept of cluster here is defined by drawing a one-to-one correspondence between the group itself and a component of the mixture. Operationally, after having estimated the model, usually via maximum likelihood by means of the EM algorithm \citep{dempster1977maximum}, the allocation is obtained via maximum \emph{a posteriori} (MAP) classification by assigning the $i$th observation to cluster $k^*$ if 
\begin{equation*} 
k^* = \argmax_k \frac{\hat{\pi}_k \varphi_k(x_i | \hat{\theta}_k)}{\sum_{k=1}^K \hat{\pi}_k \varphi_k(x_i | \hat{\theta}_k)} \; . 
\end{equation*}

When practically resorting to model-based clustering in order to obtain a partition, several choices have to be made as, for example, the number of groups $K$, the parametric specification of the mixture components or a specific parsimony-inducing parametrization of $\Sigma_k$ in the Gaussian case. Since each combination of these possibilities can be seen as a different model, it is clear how model selection steps play an essential role in this framework. Indeed several different models corresponding to such combinations are usually estimated, the best one is then chosen according to an information criterion such as the BIC \citep{schwarz1978estimating} or the ICL \citep{biernacki2000assessing} and successively used to obtain the final partition.

\begin{figure}[t]
\centering     
\subfigure{\includegraphics[height=4.9cm, width=5.5cm]{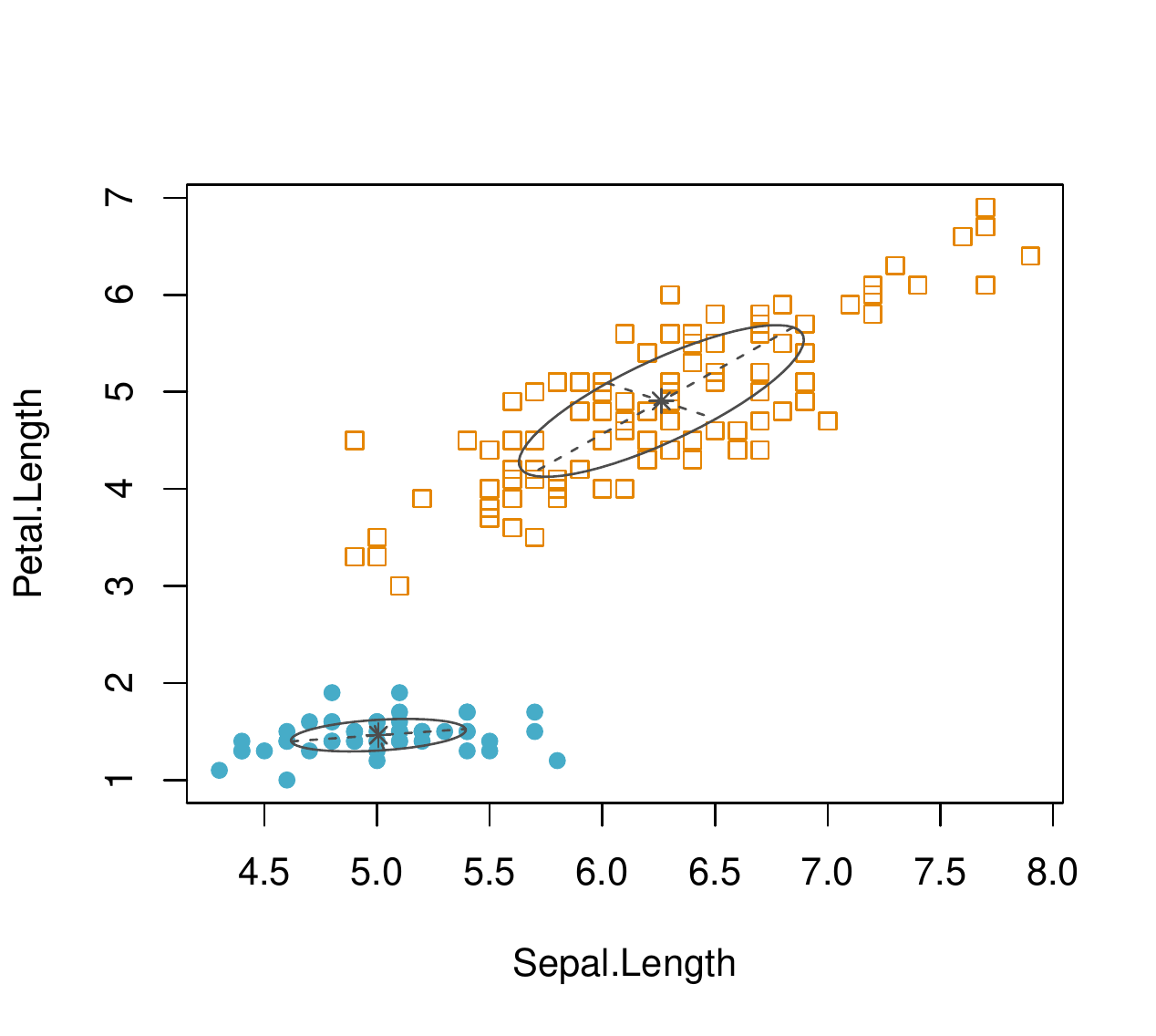}}
\subfigure{\includegraphics[height=4.6cm, width=5.7cm]{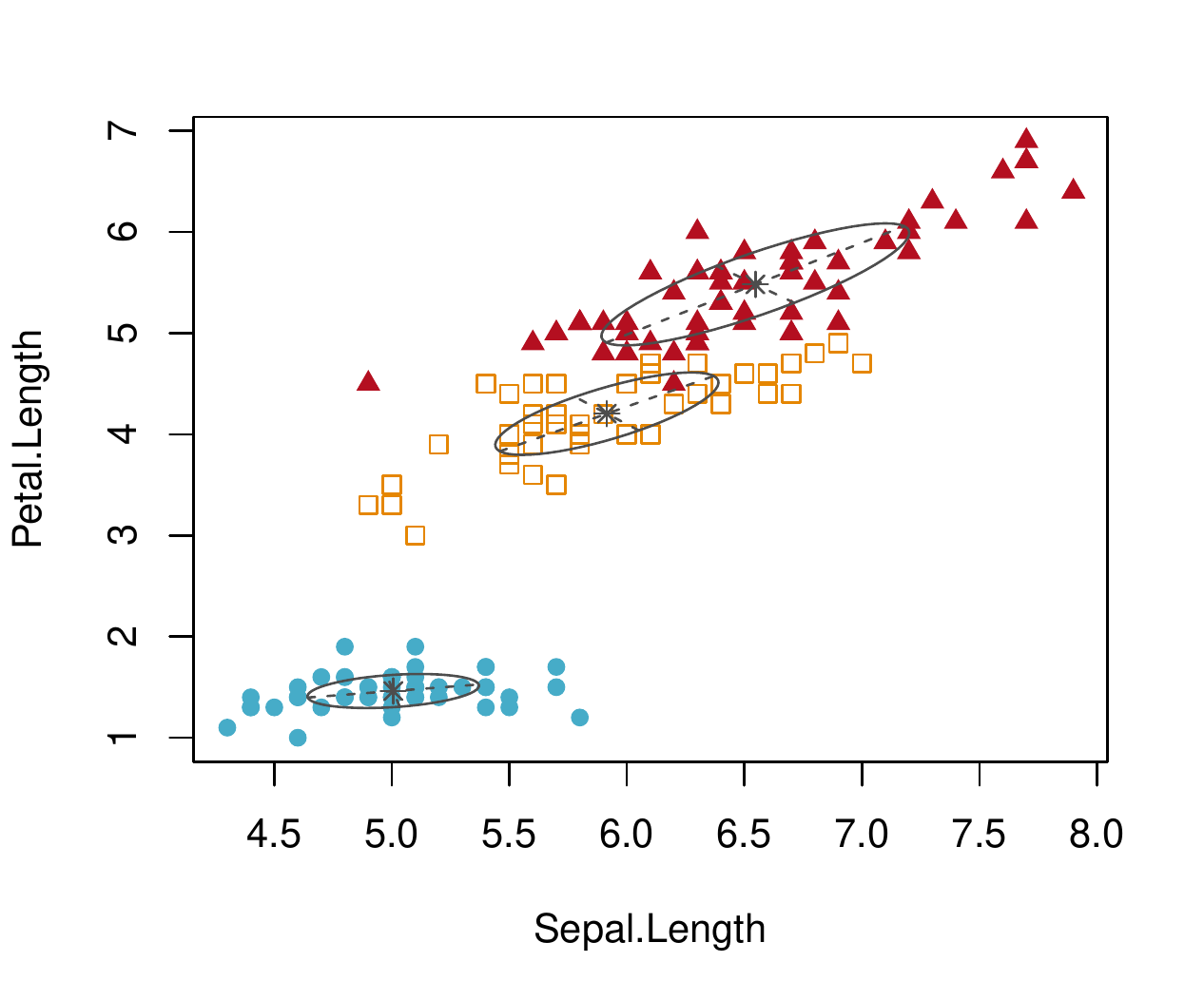}}
\caption{Example on Iris data: on the left the partition induced by the best model according to the Bayesian information criterion ($\text{BIC}=-561.72$). On the right the partition induced by the second best model ($\text{BIC}=-562.55$).}
\label{fig:irisexample}
\end{figure}

The way of proceeding, usually referred to as the \emph{single best model paradigm}, could be sub-optimal especially when differences among values of the information criterion across competing models are close. As an illustrative example we analyze the widely known Iris dataset by employing Gaussian mixture models with $K=1,\dots,9$ with all the possible parametrizations of the component covariance matrices available in \texttt{mclust} package \citep[see][for further details]{mclust}. In Figure \ref{fig:irisexample} the results obtained by the two best models according to the BIC, obtained on the four dimensional space, are shown on the subspace spanned by the variables sepal length and petal length. Even if no formal criterion is available in order to check if the difference between the values of the BIC is significant, they appear quite close. In fact, the true labels indicate the presence of three classes, here adequately captured by the second best model. Even if a proper quantification of the number of groups in these data is still controversial, since the presence of three species does not necessarily imply the presence of three groups, it seems natural to ask if, by discarding completely the second best model, useful information on the data is thrown away.

In this setting combining competitive models together may lead to a gain in robustness, stability and in the quality of the partition, as often witnessed in the supervised framework. 

In a parametric clustering framework the idea of combining different models has been developed in order to obtain partitions based on an average of models rather than on a single one. Both the works of \cite{russell2015bayesian} and \cite{wei2015mixture} propose a Bayesian model averaging approach to postprocess the results of model-based clustering. A key issue pointed out in both the proposals consists in the need of selecting an invariant quantity, i.e. a quantity having the same meaning across all the models in the ensemble, to average on. In parametric clustering this represents a cumbersome problem since the models to mix together could possibly have different number of groups as it happens, for example, if the ensemble is built with the two models in Figure \ref{fig:irisexample}; as a consequence, parameters spaces have different dimensions, thus preventing the chance to average directly parameters estimates. \cite{wei2015mixture} overcome this issue by introducing a component merging step in the procedure. Alternatively, \cite{russell2015bayesian} consider a similarity matrix as the invariant quantity, built on the agreement of cluster assignment of pairs of observations. They obtain an ensemble similarity matrix by averaging the candidate models ones. Afterwards the resulting matrix, where the $(i,j)$\emph{th} entry represents the averaged probability of $x_i$ and $x_j$ to belong to the same cluster, is considered to obtain partitions adopting a hierarchical clustering approach. 

In this work we take a different path with respect to the ones mentioned above. The issue is tackled directly at its roots, by exploiting the essential role assumed by the density in the considered framework. Therefore, recasting the problem to a density estimation one, the density itself is chosen as the invariant quantity to be averaged. Note that some promising results, from a density estimation perspective only, have been obtained by \citet{glodek2013ensemble} whose work shares some similarities with our proposal.\\
Let $\{f_m(\cdot | \hat{\Psi}_m) \}_{m=1,\dots,M}$ be a set of estimated candidate mixture models. In this Section the number $M$ of models to average is considered as given, and we refer the reader to Section \ref{sec:chscrucca_discussion} for a discussion about this aspect. In the rest of the work we focus specifically on mixtures of Gaussian densities. This choice is not binding for subsequent developments since, in principle, the ensemble may be populated by mixture models with different parametric specifications for the component densities. 
A new estimator, being a convex linear combination of the estimated densities $f_m(\cdot|\hat{\Psi}_m)$, is introduced: 
\begin{eqnarray}\label{eq:eq1}
\tilde{f}(x; \alpha) = \sum_{m=1}^M \alpha_m f_m(x | \hat{\Psi}_m) \; ,
\end{eqnarray}
with $\alpha_m>0$, $\sum_m \alpha_m = 1$, representing the weight assigned to the $m$th model for all $m = 1,\dots,M$. 
A key aspect, as it will be discussed in Section \ref{sec:chscrucca_estimation}, consists in properly estimating the model weights in order to guarantee that models describing more adequately the underlying density will count more in the resulting estimator. 

The rationale behind our proposal draws strength from some results obtained by \cite{rigollet2007linear}. Here the authors show that, under some fairly general regularity assumptions, linearly aggregating density estimators leads asymptotically to an improvement in the resulting density under $L_2$-loss perspective.  Hence, by possibly improving the quality of the density estimates, we aim at obtaining better characterizations of the relevant patterns in the data, leading to more refined partitions. 

Even if the estimator (\ref{eq:eq1}) is still a mixture model we cannot obtain a partition as usually carried out in parametric clustering, thus resorting to the one-to-one correspondence among groups and components. As an illustrative example, consider an ensemble formed by two mixture models, with two and three components respectively. In this situation $\tilde{f}(\cdot;\alpha)$ will result in a five component mixture model hence giving contradictory indications about the number of groups with respect to the models that have been mixed together. This issue shares strong contact points with the situations where the number of components exceeds the number of groups; for example a two components Gaussian mixture may result in a unimodal density leading to a counterintuitive partition with no clear separation between the two groups. In the model-based clustering framework the problem has been addressed by resorting to merging procedures \citep{baudry2010combining,hennig2010methods} where mixture components are combined together and their union seen as a single cluster. 

In this work we take a different path and naturally circumvent the problem by shifting the concept of cluster to its modal formulation.
Consistently, the grouping structure is searched in the modality of an estimate of the density (\ref{eq:eq1}). Hence, each cluster is built by gathering all those observations ascending to the same mode of the density. Operationally, a partition is obtained by means of a mode-searching algorithm. Our proposal shares some conceptual connections with the work by \citet{hennig2010methods} where the author proposes merging methods aimed at finding unimodal clusters. 

The proposed solution, staying in the realm of density-based clustering, inherits and enjoys its relevant strengths as the chance to frame the problem in a standard inferential setting where proper statistical tools can be employed for evaluation, and to obtain whole sample space partitions whose features are inferentially explorable. Moreover it has been shown already \citep[see][]{scrucca2016identifying,chacon2016mixture} that blending together parametric and nonparametric approaches to clustering can lead to some relevant improvements in some, otherwise troublesome, situations.

\subsection{Model estimation}\label{sec:chscrucca_estimation}
The procedure outlined in Section \ref{sec:chscrucca_frameworkmodelspec} requires a practical way to estimate the density in (\ref{eq:eq1}). Note that, since $\hat{\Psi}_m$ has been previously estimated, the only unknown parameters involved are the $\alpha_m$s. These parameters represent the weights to be assigned at every single model in the ensemble, hence their estimation is crucial in governing the resulting shape of the density, its modal structure and consequently the final partition. A reasonable estimation procedure would result in giving nearly zero weights to those models in the ensemble which do not suitably capture the features of the underlying density, while weighting more the adequate ones. 

In order to obtain an estimate for the weight vector $\alpha=(\alpha_1,\dots,\alpha_M)$, based on the sample $\mathcal{X}$, we aim to maximize the log-likelihood of the model (\ref{eq:eq1}), defined as
\begin{eqnarray}\label{eq:loglik}
\ell(\alpha;\mathcal{X}) = \sum_{i=1}^n\log \sum_{m=1}^M \alpha_m f_m(x_i|\hat{\Psi}_m). 
\end{eqnarray}
However, if (\ref{eq:loglik}) is considered as the objective function to maximize, the procedure will incur in the overfitting problem since the most complex models in the ensemble, which provide a better fit by construction, will weight more. This behaviour will commonly result in wiggler estimates not appropriately seizing the relevant features of the density, hence some regularization has to be considered in the estimation. 

A tentative solution has been proposed by \cite{smyth1999linearly} where a \emph{stacking} procedure is adapted to the density estimation framework. The authors avoid to fall into the overfitting trap by exploiting a cross-validation scheme when combining the candidate models to obtain ensemble density estimates.

We take a different path by replacing the log-likelihood in (\ref{eq:loglik}) with a penalized version, generally defined as
\begin{eqnarray}\label{eq:penlik}
\ell_P(\alpha;\mathcal{X}) = \ell(\alpha;\mathcal{X}) - \lambda g(\alpha, \nu) \; .
\end{eqnarray}
Here $g(\cdot)$ is a penalty function to be specified, $\nu = (\nu_1,\dots, \nu_M)$ is a vector measuring the complexity of the models in the ensemble, while $\lambda$ is a parameter controlling for the strength of the penalization. Within this general framework, we set $\nu_m$ to be the cardinality of $\hat{\Psi}_m$, as it appears a sensible proxy of the complexity of the $m$th model. Additionally, we consider $g(\alpha,\nu)=\sum_m \alpha_m\nu_m$ as a simple choice which guarantees a stronger penalization to the most complex models. Note that, since both $\alpha_m$ and $\nu_m$ are positive numbers, $\sum_m \alpha_m\nu_m = \sum_m |\alpha_m\nu_m|$. As a consequence our penalty might be seen as a generalization of the LASSO one where $\nu_m$ is introduced to account for the complexities of the models in the ensemble.

For a given value of $\lambda$, the parameters is then estimated to maximize the penalized log-likelihood
$$
\hat{\alpha} = \hat{\alpha}(\lambda) = \argmax_\alpha \ell_P(\alpha; \mathcal X). 
$$
To this end, due to the mixture structure easily recognizable in (\ref{eq:eq1}), we can resort to a slightly simplified version of the EM-algorithm in order to maximize the penalized log-likelihood (\ref{eq:penlik}); note that the number of free parameters to estimate is equal to $M-1$ since $\hat{\Psi}_m$ with $m=1,\dots, M$ have to be considered as fixed and $\sum_{m=1}^M \alpha_m=1$. At iteration $t$ of the \emph{E-step}, conditionally to an estimate $\hat{\alpha}^{(t-1)}$ for the vector $\alpha$ at the previous iteration, we compute
\begin{equation}
\tau_{mi}^{(t)} = \frac{\hat{\alpha}_m^{(t-1)} f_m(x_i | \hat{\Psi}_m)}{\sum_{m'=1}^M \hat{\alpha}_{m'}^{(t-1)}f_{m'}(x_i | \hat{\Psi}_{m'})} \; .
\end{equation}

\noindent Then the \emph{M-step} will consist in maximizing, with respect to $\alpha$, the expected value of the complete-data penalized log-likelihood, in our setting expressed as 

\begin{equation}\label{eq:qfunct}
Q_p(\alpha; \hat{\alpha}^{(t-1)}) = \sum_{m=1}^M\sum_{i=1}^n \tau_{mi}^{(t)}[\log\alpha_m + \log f_m(x_i | \hat{\Psi}_m)] - \lambda \sum_{m=1}^M \alpha_m\nu_m \;  , 
\end{equation}
under the constraint $\sum_{m} \alpha_m =1$ with $\alpha_m > 0, \; \forall m = 1, \dots, M$. Since closed form solutions are not available, $\hat{\alpha}^{(t)}$ is obtained by maximizing (\ref{eq:qfunct}) numerically. As usual, the two steps will be iterated until a convergence criterion is met. Several initialization strategies might be adopted in order to obtain $\hat{\alpha}^{(0)}$: in this work we consider a uniform initialization where, in the first step of the EM-algorithm, all the models in the ensemble are equally weighted.

Regarding the choice of $\lambda$, some more caution is needed, since an accurate selection turns out to be essential in order to obtain a meaningful estimate which properly reflects the geometrical structure of the underlying density. In this work some different options have been taken into consideration. A first, possible, strategy consists in estimating $\lambda$ by means of the observed data resorting to a cross-validation scheme defined as follows:
\begin{itemize}
  \item Randomly split the set $\{ 1,\dots, n\}$ into $V$ equally-sized subsets $\mathcal{F}_1,\dots,\mathcal{F}_V$; 
  \item For $v=1,\dots,V$:
  \begin{itemize}
    \item Consider as a training sample $\mathcal{X}_{\text{train}}(v) = \{ x_i \}_{i \notin \mathcal{F}_v}$ and as a test sample $\mathcal{X}_{\text{test}}(v) = \{ x_i \}_{i \in \mathcal{F}_v}$;
    \item For varying $\lambda$ in a reasonable grid $\Lambda$, maximize $\ell_P(\alpha;\mathcal{X}_{\text{train}}(v))$ and obtain $\hat{\alpha}_v(\lambda)$;
    \item For each $x \in \mathcal{X}_{\text{test}}(v)$ predict the density $\tilde{f}(x;\hat{\alpha}_v(\lambda))$;
  \end{itemize}
  \item Define a \emph{test log-likelihood}
  $$
  \ell_{\text{test}}(\lambda) = \sum_{v=1}^V \sum_{x \in \mathcal{X}_{\text{test}}(v)} \log \tilde{f}(x;\hat{\alpha}_v(\lambda))
  $$
  and select 
  $$
  \lambda_{CV} = \argmax_{\lambda \in \Lambda} \ell_{\text{test}}(\lambda)
  $$
  The selected $\lambda_{CV}$ is finally used to obtain an estimate of $\alpha$ based on the whole sample. 
\end{itemize}

Another reasonable approach consists in taking inspiration from the formulations of some information criteria which may be seen, in all respects, as penalized likelihood. Therefore we introduce the \emph{AIC-type} and the \emph{BIC-type} penalizations, stemming directly from the definitions of AIC and BIC, that induce penalized log-likelihoods defined as
\begin{eqnarray}\label{eq:penloglikaicbic}
\ell_{P,AIC}(\alpha; \mathcal{X}) &=& 2\ell(\alpha;\mathcal{X}) - 2\sum_{m=1}^M \alpha_m \nu_m \\
\ell_{P,BIC}(\alpha;\mathcal{X}) &=& 2\ell(\alpha; \mathcal{X}) - \log(n)\sum_{m=1}^M \alpha_m \nu_m \; ,
\end{eqnarray}
hence implying $\lambda_{AIC}=1$ and $\lambda_{BIC}=\log(n)/2$ according to the formulation in (\ref{eq:penlik}).

\noindent Although requiring an higher computational effort, it stands to reasons that the cross-validation based approach has some relevant advantages in the regularization process. By resorting to a fully data-driven selection of $\lambda$, we end up with a more adaptive parameter than $\lambda_{\text{BIC}}$ and $\lambda_{\text{AIC}}$, both in terms of sample size and features of the observed data. However, the latter penalties are computationally faster and simple rules of thumbs enable, in practice, to produce satisfactory results, as will be discussed in Section 4. 

Once the density (\ref{eq:eq1}) is estimated, a partition is operationally obtained by identifying its modal regions. To this aim we consider the so called Modal EM \citep[MEM;][]{li2007nonparametric} in the modified version proposed by \citet{scrucca2020fast}. Designed for densities which are built as mixtures, this technique alternates two iterative steps in the guise of the EM, but unlike the EM its goal is to find the local maxima of the density.
Since the density (\ref{eq:eq1}) may still be seen as a peculiar mixture model, MEM can be fruitfully adapted to our situation where, given an estimate $\hat{\alpha}$, a starting value $x^{(0)}$ and setting initially $r=0$, the iterative steps are defined as follows
\begin{itemize}
  \item Let $$p_m=\frac{\hat{\alpha}_m f_m(x^{(r)}|\hat{\Psi}_m)}{\sum_{m'=1}^M \hat{\alpha}_{m'}f_{m'}(x^{(r)} | \hat{\Psi}_{m'})}$$
  \item Update $$ x^{(r+1)} = \argmax_x \sum_{m=1}^M p_m \log f_m(x|\hat{\Psi}_m) \; . $$
\end{itemize}
The algorithm iteratively performs these steps until a convergence criterion is met. The outlined iterative procedure draws a path leading to a local maximum of the density \citep[see][for a proof of the ascending property of the algorithm]{li2007nonparametric}. Lastly, a partition is operationally obtained by using each observation $\{x_i\}_{i=1,\dots,n}$ in the sample as an initial value in the MEM and by grouping together those observations converging to the same mode.

\section{Some remarks}\label{sec:chscrucca_discussion}

In this section we discuss further the procedure introduced so far by pointing out some practical considerations and highlighting its properties along with some links with other existing methods. 

\begin{remark}\label{remark:chscrucca_rk2}
Estimator (\ref{eq:eq1}) has been introduced by considering the models to be mixed $f_m(\cdot | \hat{\Psi}_m)$, as well as their number $M$, as given. In fact, a virtually huge number of models could be estimated, and selecting which ones should enter in the ensemble could have some impact on the resulting partitions. In practice, when choosing the ensemble size, different paths might be considered. 

A first possible approach consists in populating the ensemble with all the models estimated in the previous step of the procedure, being reasonable candidates and representing a wide batch of alternatives recording a general uncertainty. In such a way the possibly troublesome selection of $M$ is somehow circumvented by letting the penalty term to do the job. In practical applications the penalized estimation strategy would indeed shrink towards zero the weights of the models considered as irrelevant hence somehow automatically selecting $M$, here defined as the number of models considerably weighted in the ensemble. 

Another alternative may consist in in considering an \emph{Occam's window} to choose a set of models as proposed by \cite{madigan1994model}. The main idea is to discard those models providing estimates being qualitatively too far from the ones provided by the best model. A rule of thumb would be to discard the $m$th model if $|\text{BIC}_{\text{best}}- \text{BIC}_{m}| > 10$, where $\text{BIC}_{\text{best}}$ and $\text{BIC}_{m}$ represent respectively the values of the BIC for the best model and for the $m$th one. Lastly another viable approach consists in choosing $M$ subjectively and picking those models, among the estimated ones, resulting in a good fitting of the data. In this case $M$ should vary also reflecting the case-specific uncertainty witnessed in the modelling process.

Finally a word of caution; finding substantial arguments that motivate some general recommendations is challenging and cannot leave aside the specificities of the data and of the problem at hand.
\end{remark}

\begin{remark}\label{remark:chscrucca_rk4}
The estimation procedure outlined in Section \ref{sec:chscrucca_estimation} is fully frequentist in nature. Alternatively, a Bayesian approach could be an interesting development claiming some advantages. The work by \cite{malsiner2017identifying} faces, from a Bayesian perspective, the estimation of mixtures of mixture models. Even if the underlying motivation is different some ideas could be fruitfully borrowed and exploited in order to average different mixture models. As an example, the consideration of a shrinkage prior on the weights of the models in the ensemble could practically overcome the previously discussed issue of selecting $M$.
\end{remark}

\begin{remark}\label{remark:chscrucca_rk3}
Model selection often precedes inference that is usually conducted considering the chosen model as fixed. However, since the selection is itself data-dependent, it possesses its own variability. Drawing inference without accounting for the selection of the model corresponds to neglect completely a source of uncertainty and usually results in anti-conservative statements \citep{leeb2005model}. Even in the full awareness of the fact that, in parametric clustering, the main focus tipically lies on obtaining partitions rather than on inference or uncertainty quantification, we believe that a model averaging approach can entail better estimation properties and more informative confidence intervals for the parameters when needed.
\end{remark}

\begin{remark}\label{remark:chscrucca_rk1}
In the supervised framework ensemble approaches have been found tremendously effective in improving predictions of a plethora of different models. For those techniques it has been frequently noticed  \citep[see, e.g.][]{dietterich2000experimental} how the concept of \emph{diversity} is a key factor in increasing classification performances of the \emph{base learners} that are combined. As a consequence, often  weak learners are considered in the supervised context. These classifiers are highly unstable, consequently different one from the others, as they possibly focus on distinct features of the observed data. Even in a clustering framework the impact of the diversity among the combined partitions has been empirically studied and proved to be impactful by \cite{fern2003random} and \cite{kuncheva2004using}. \\ We are aware that, when the proposed method is used to go beyond the \emph{single best model} paradigm, the models in the ensemble cannot be considered as weak and consequently diversity among them is not achieved. Nonetheless, even if introduced with a specific aim, the proposal can in principle be exploited in all those cases where averaging multiple density-induced clusterings could be fruitful. As a consequence, the diversity can be somehow determined for example averaging densities computed on bootstrap samples or on general subsamples of the observed data. Since initialization plays a crucial role when resorting to the EM algorithm  \citep[see, e.g.][]{scrucca2015improved}, another appealing application consists in combining models estimated using different starting values. As a consequence of the estimation instability these models would probably be more heterogeneous hence entailing greater diversity. 
\end{remark}

\begin{remark}\label{remark:chscrucca_rk6}
The model introduced so far, despite being based on a different rationale, shares some connections with the general framework of \emph{Deep Gaussian Mixture Models} investigated by \citet{viroli2019deep}. Deep Gaussian Mixture Models are networks of multiple layers of latent variables distributed as a mixture of Gaussian densities. Since the outlined representation encompasses the specification of a mixture of mixtures \citep{li2005clustering}, model (\ref{eq:eq1}) can be seen as a two layers Deep Gaussian Mixture Model where the parameters involved in the inner layer are fixed.
\end{remark}


\section{Empirical analysis}\label{sec:chscrucca_results}
\subsection{Synthetic data}\label{sec:chscrucca_simulation}
\noindent The idea of averaging together different densities to obtain a more informative summary for clustering purposes is explored in this section via simulations. The simulation study has multiple goals. On one side we want to evaluate the performances of our proposal in terms of the quality of the produced density estimates. These performances are studied with respect to the true and known density function considering the MISE as evaluating criterion. On the other hand the clustering performances of the proposed method are investigated. As an assessment criterion we employ the \emph{Adjusted Rand Index} \citep[ARI,][]{hubert1985comparing} between the obtained partitions and the true component memberships of the observations. An additional aim consists in evaluating how the sample size impacts on these comparisons.  

As a side goal of the numerical explorations we want to study which penalization strategy introduced in Section \ref{sec:chscrucca_estimation} produces more satisfactory results. In particular, we evaluate whether the increased computational costs implied by the cross-validation are worth the effort or if less intensive strategies such as the \emph{BIC-} and \emph{AIC-type} penalties produce comparable results. Lastly, we want to compare our proposals with some reasonable competitors. We consider a fully parametric approach, using the single best model chosen among a set of Gaussian mixture models corrisponding to combinations of the number of components and of different covariance matrix parametrizations.
Moreover, we consider a nonparametric clustering method where the density is estimated using a kernel estimator with unconstrained gradient as bandwidth matrix, a standard choice in nonparametric density literature  \citep[see][for a detailed tractation]{chacon2018multivariate}. The partition is afterwards practically obtained resorting to the \emph{mean-shift algorithm} \citep{fukunaga1975estimation,cheng1995mean}. Furthermore, we examine also an hybrid approach consisting in finding the modes, via Modal EM algorithm, of the density estimated by the single best model. The possible improvements introduced by our proposal may be due to two different motivations: the first related to a better estimation of the underlying density while the second is concerned with the modal-inspired allocation procedure. Considering an hybrid approach as a competitor can help disentangling properly these distinct sources.

\begin{figure}[t!]
\center
\includegraphics[scale=0.65]{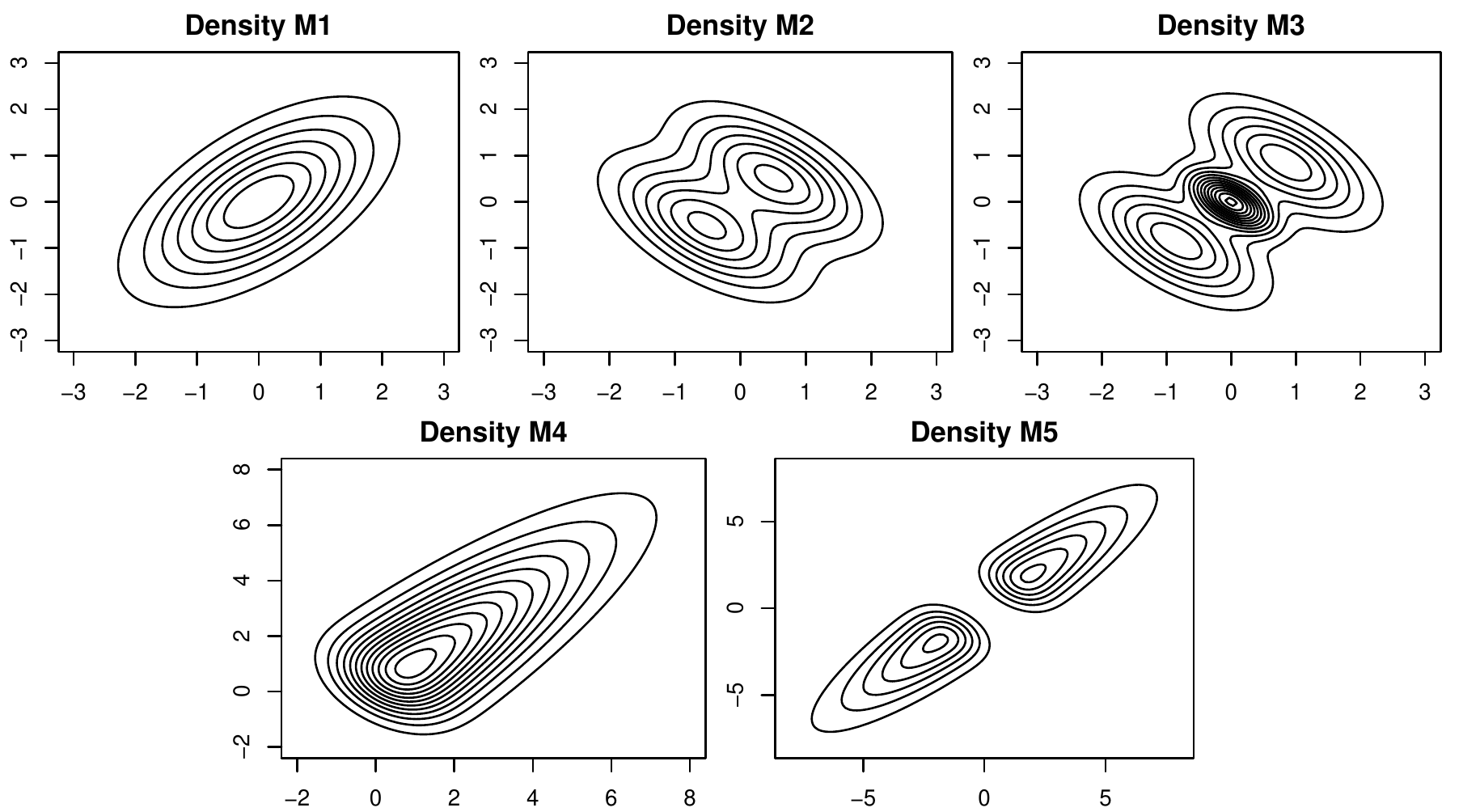}
\caption{Bivariate density functions used in the simulation study.}
\label{fig:sim_densities_scrucca}
\end{figure}

A total of $B=200$ samples have been drawn, with sizes $n \in \{ 500, 5000 \}$, for each of the bivariate densities depicted in Figure \ref{fig:sim_densities_scrucca} and whose parameters are reported in Appendix \ref{App:settings}. These densities have been considered to encompass different situations which pose different challenges from a model-based clustering perspective. The densities on the top panels of Figure \ref{fig:sim_densities_scrucca} represent indeed settings where the single best model is expected to achieve satisfactory results, being the data generated from Gaussian mixtures. On the other hand the densities on the bottom panels, showing strong asymmetric behaviours, constitute more challenging settings where Gaussian mixture models generally produce inadequate partitions. All the reported analyses have been conducted in the \texttt{R} environment \citep{rcore} with the aid of the \texttt{mclust} \citep{mclust}, \texttt{ks} \citep{kspack} and \texttt{EMMIXskew} \citep{emmixskew} packages.

Throughout the simulations we estimated a total of 126 models, corresponding to the default setting in \texttt{mclust}, where the 14 different parametrizations of the component covariance matrices \citep[see][for more details]{celeux1995gaussian,mclust} are combined with varying number of mixture components $K=1,\dots,9$. Afterwards we have considered $M=30$ best models ranked according to their BIC values, coherently with Remark \ref{remark:chscrucca_rk2} in Section \ref{sec:chscrucca_discussion}; this choice moves towards the direction of retaining a large number of models, letting the estimation procedure to select the most relevant ones, while keeping the computations feasible. Note that some additional analyses have shown that clustering performances are not strongly influenced by a further increase in the ensemble size. Moreover the choice of the information criterion to rank the models and to select the best $M$ ones is neither constraining nor strongly influential; here the BIC has been considered in order to be consistent with the standard practice in the model-based clustering framework. We also explored the option of selecting $M$ by the \emph{Occam's window} to build the ensemble as discussed in Remark \ref{remark:chscrucca_rk2}; nonetheless results, not reported here, indicate that this strategy often leads to the selection of a too small subset of models due to the strong reliance on BIC. The three options $\lambda_{AIC}$, $\lambda_{BIC}$ and $\lambda_{CV}$ discussed in Section \ref{sec:chscrucca_estimation} are evaluated, the last one resorting to a $V$-fold cross-validation scheme with $V=5$.
 
\begin{table}[t!]
\begin{center}
\begin{tabular}{lcc|cc}
\hline
 & \multicolumn{2}{c|}{$n=500$} & \multicolumn{2}{c}{$n=5000$} \\
 \hline
 & \multicolumn{2}{c|}{\includegraphics[scale=0.23]{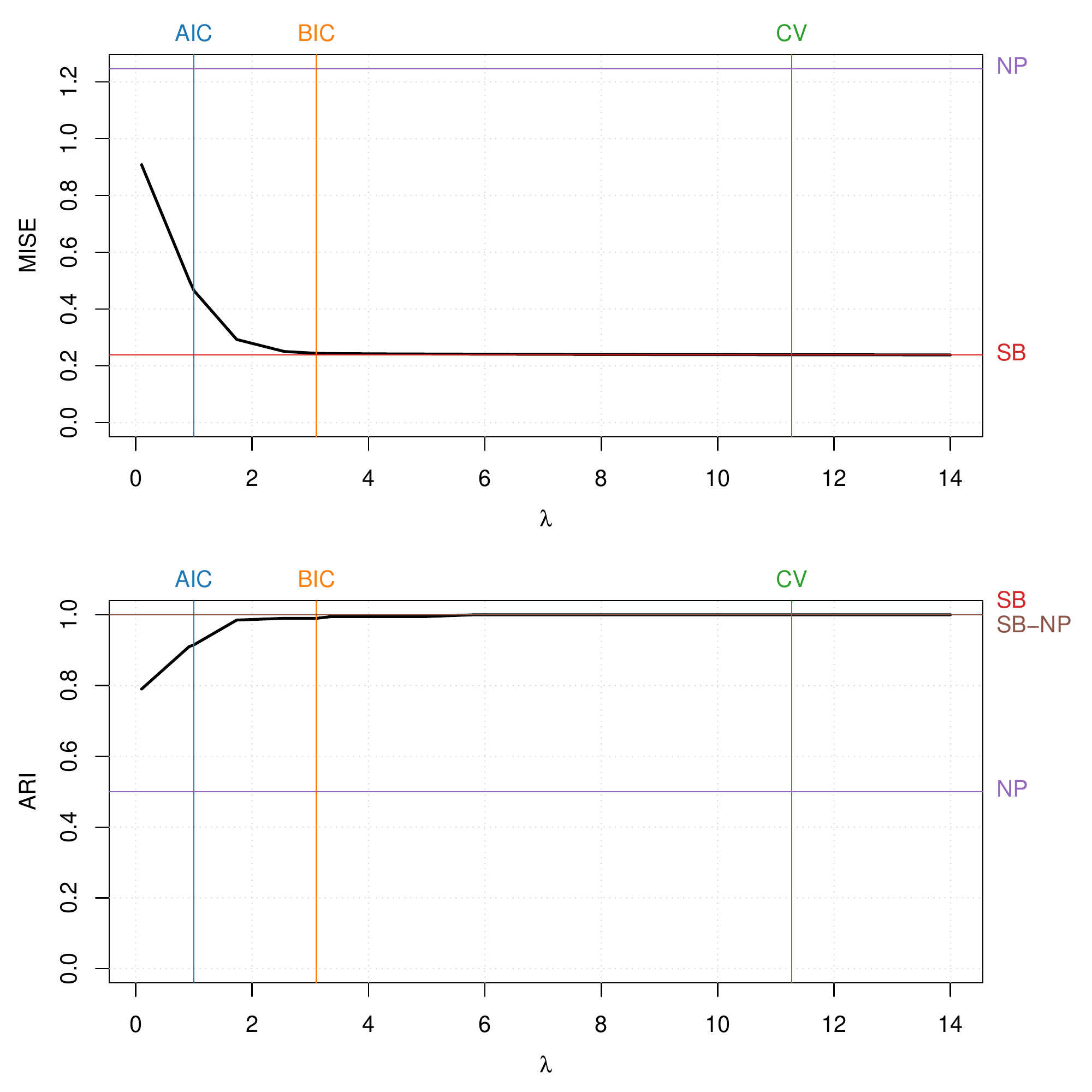}} & \multicolumn{2}{c}{\includegraphics[scale=0.23]{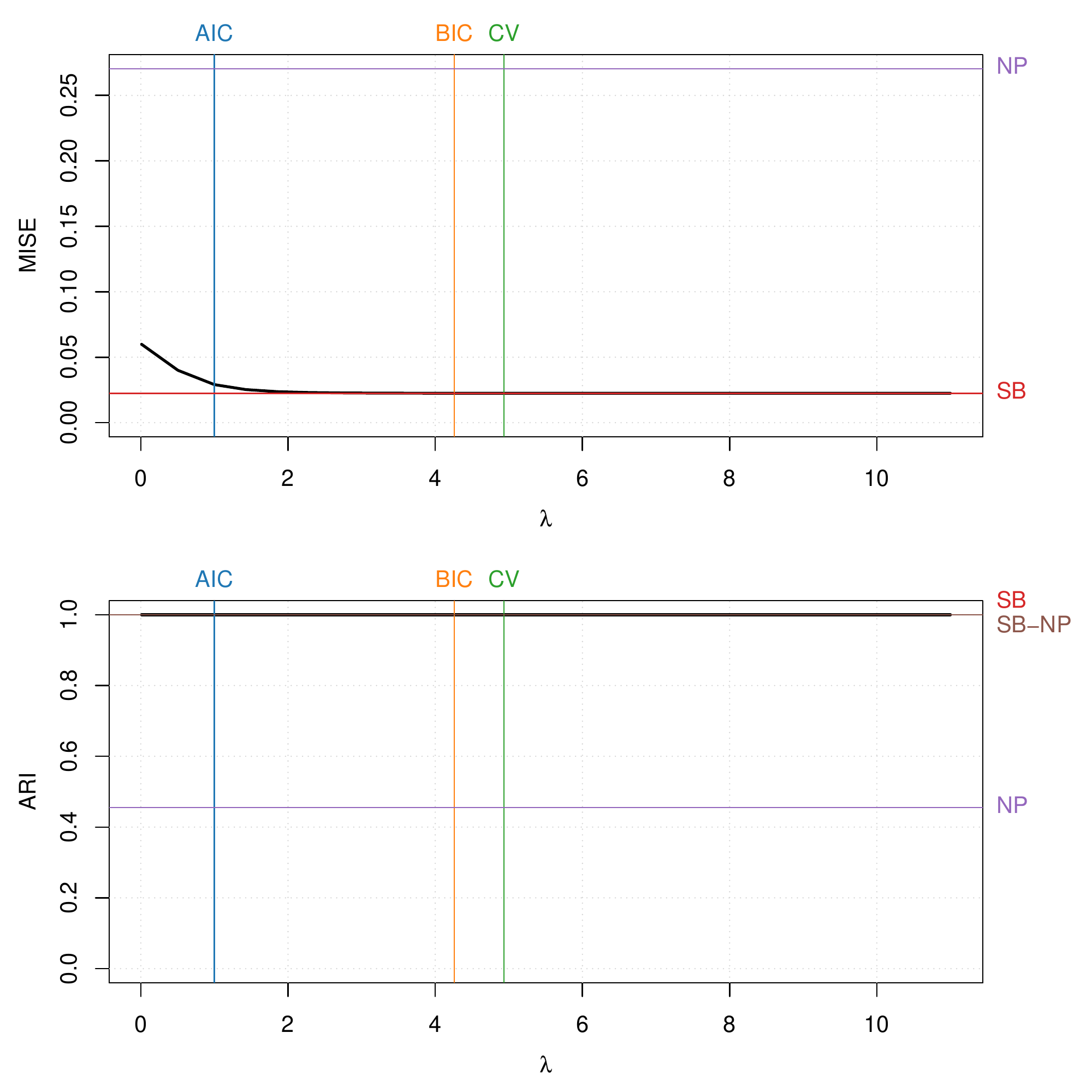}} \\ 
 \hline 
  & MISE & ARI & MISE & ARI \\ \hline
 SB & 0.238 (0.162) & 1.000 (0.000) & 0.022 (0.015) & 1.000 (0.000) \\
 NP & 1.246 (0.392) & 0.500 (0.501) & 0.270 (0.076) & 0.455 (0.499)\\
 SB-NP & - & 1.000 (0.000) & - & 1.000 (0.000) \\
 $\lambda_{CV}$ & 0.241 (0.164) & 0.995 (0.0071) & 0.024 (0.016) & 1.000 (0.000) \\
 $\lambda_{AIC}$ & 0.465 (0.380) & 0.915 (0.280) & 0.029 (0.019) & 1.000 (0.000) \\
 $\lambda_{BIC}$ & 0.244 (0.167) & 0.990 (0.099) & 0.022 (0.015) & 1.000 (0.000) \\
 \hline

\end{tabular}
\end{center}
\caption{Top panel: the MISE ($\times \, 1000$) and the ARI (black lines) as functions of $\lambda$ for $n=500,5000$. Red, purple and brown horizontal lines represent the same quantities respectively for the single best model (SB), the nonparametric approach (NP) and the hybrid approach (SB-NP). The vertical lines represent the mean values over the $B$ samples of $\lambda_{CV}$ (in green), $\lambda_{AIC}$ (in light blue) and $\lambda_{BIC}$ (in orange). Bottom panel: numerical values of the MISE ($\times \, 1000$) and average ARI (and their standard errors) over the Monte Carlo samples for the competing considered methods. Results refer to density M1.}
\label{tab:M1_scrucca}
\end{table}

\begin{table}[t!]
\begin{center}
\begin{tabular}{lcc|cc}
\hline
 & \multicolumn{2}{c|}{$n=500$} & \multicolumn{2}{c}{$n=5000$} \\
 \hline
 & \multicolumn{2}{c|}{\includegraphics[scale=0.23]{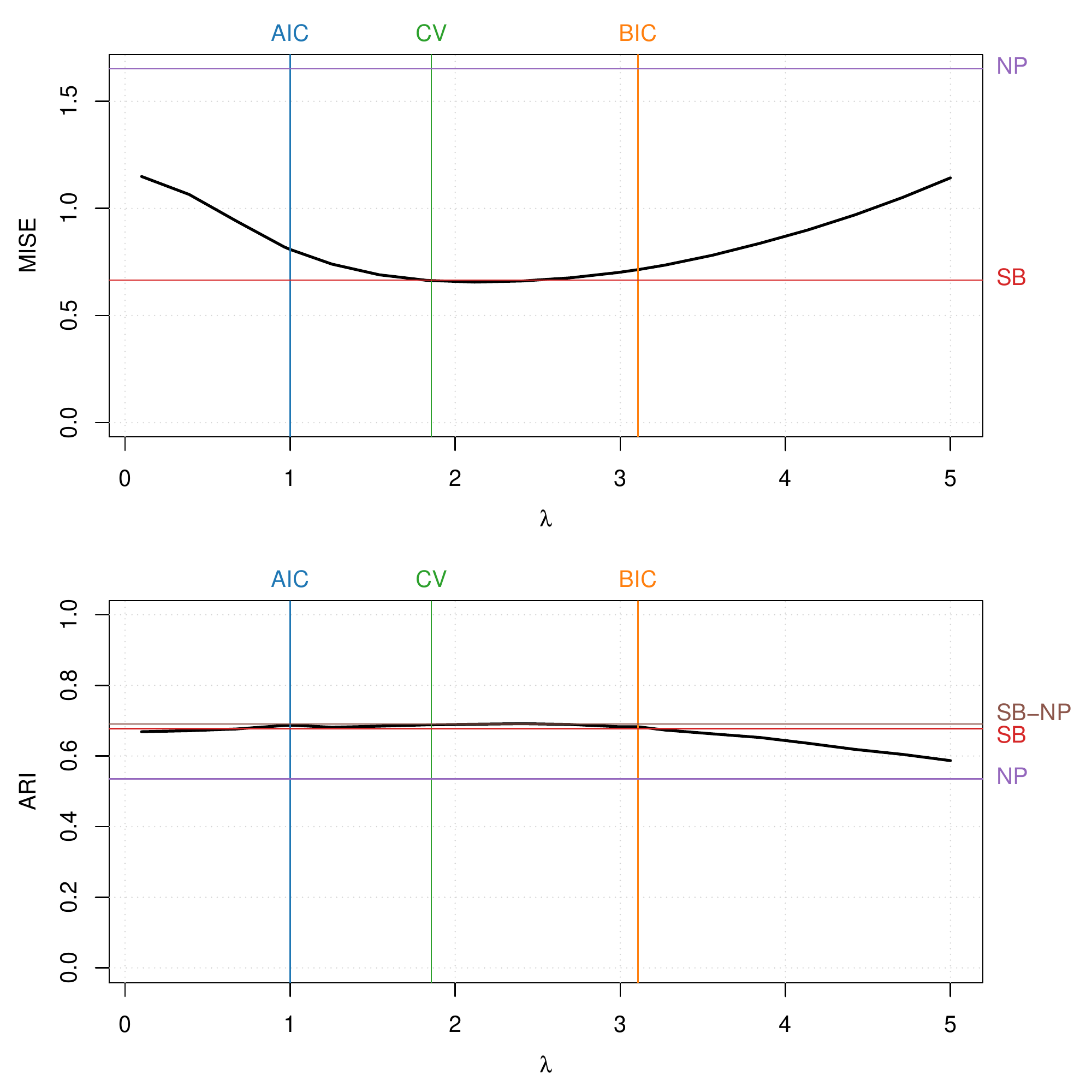}} & \multicolumn{2}{c}{\includegraphics[scale=0.23]{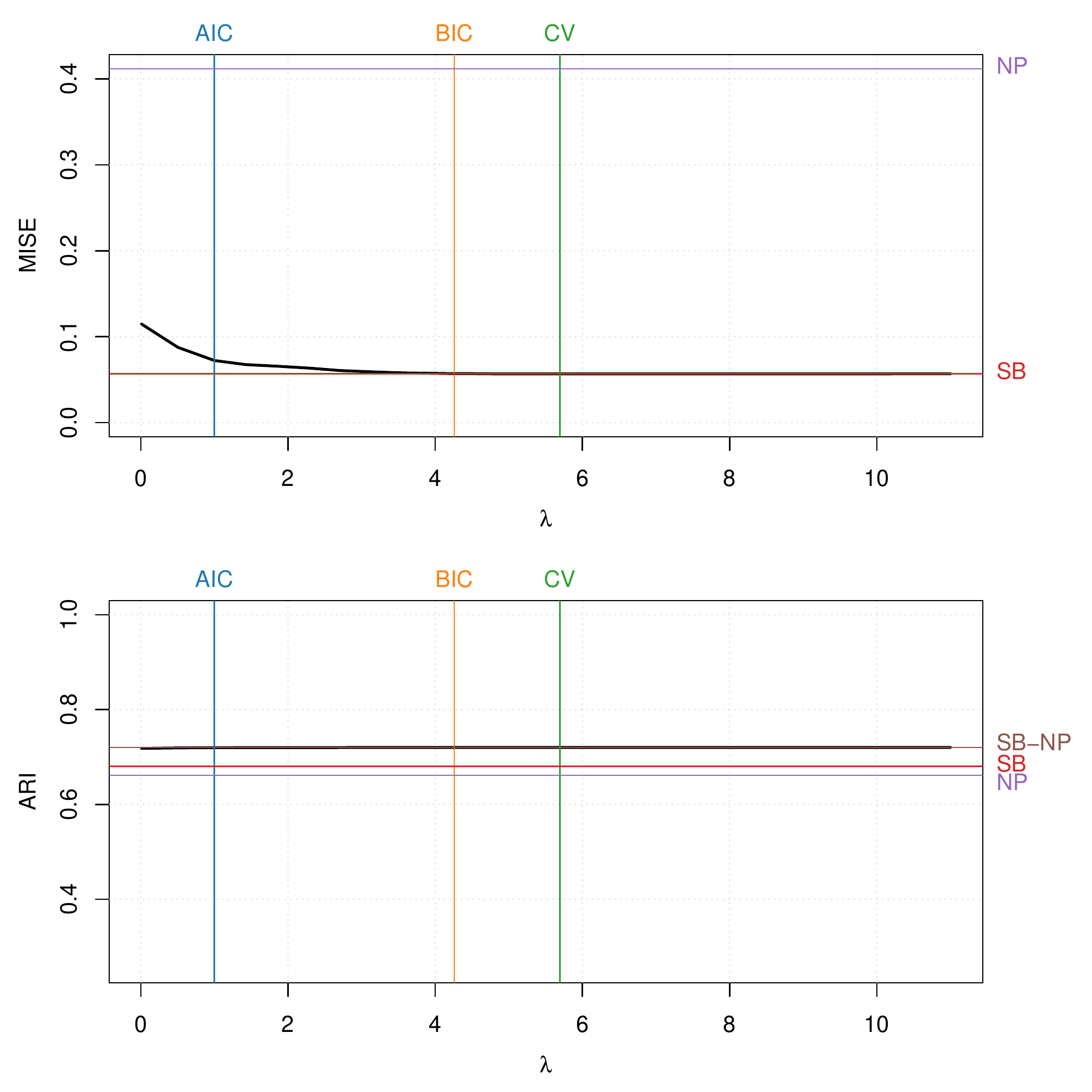}} \\ 
 \hline 
  & MISE & ARI & MISE & ARI \\ \hline
 SB & 0.666 (0.714) & 0.677 (0.125) & 0.057 (0.035) & 0.680 (0.067) \\
 NP & 1.652 (0.434)  & 0.535 (0.186) & 0.412 (0.094) & 0.661 (0.083) \\
 SB-NP & - & 0.690 (0.119) & - & 0.720 (0.012) \\
  $\lambda_{CV}$ & 0.687 (0.402) & 0.688 (0.064) & 0.058 (0.036) & 0.720 (0.012) \\
 $\lambda_{AIC}$ & 0.809 (0.435) & 0.687 (0.063) & 0.072 (0.044) & 0.719 (0.013) \\
 $\lambda_{BIC}$ & 0.714 (0.522) & 0.683 (0.129) & 0.057 (0.035) & 0.720 (0.012) \\
 \hline

\end{tabular}
\end{center}
\caption{Cf. Table \ref{tab:M1_scrucca}. Results refer to density M2}
\label{tab:M2_scrucca}
\end{table}

\begin{table}[t!]
\begin{center}
\begin{tabular}{lcc|cc}
\hline
 & \multicolumn{2}{c|}{$n=500$} & \multicolumn{2}{c}{$n=5000$} \\
 \hline
 & \multicolumn{2}{c|}{\includegraphics[scale=0.23]{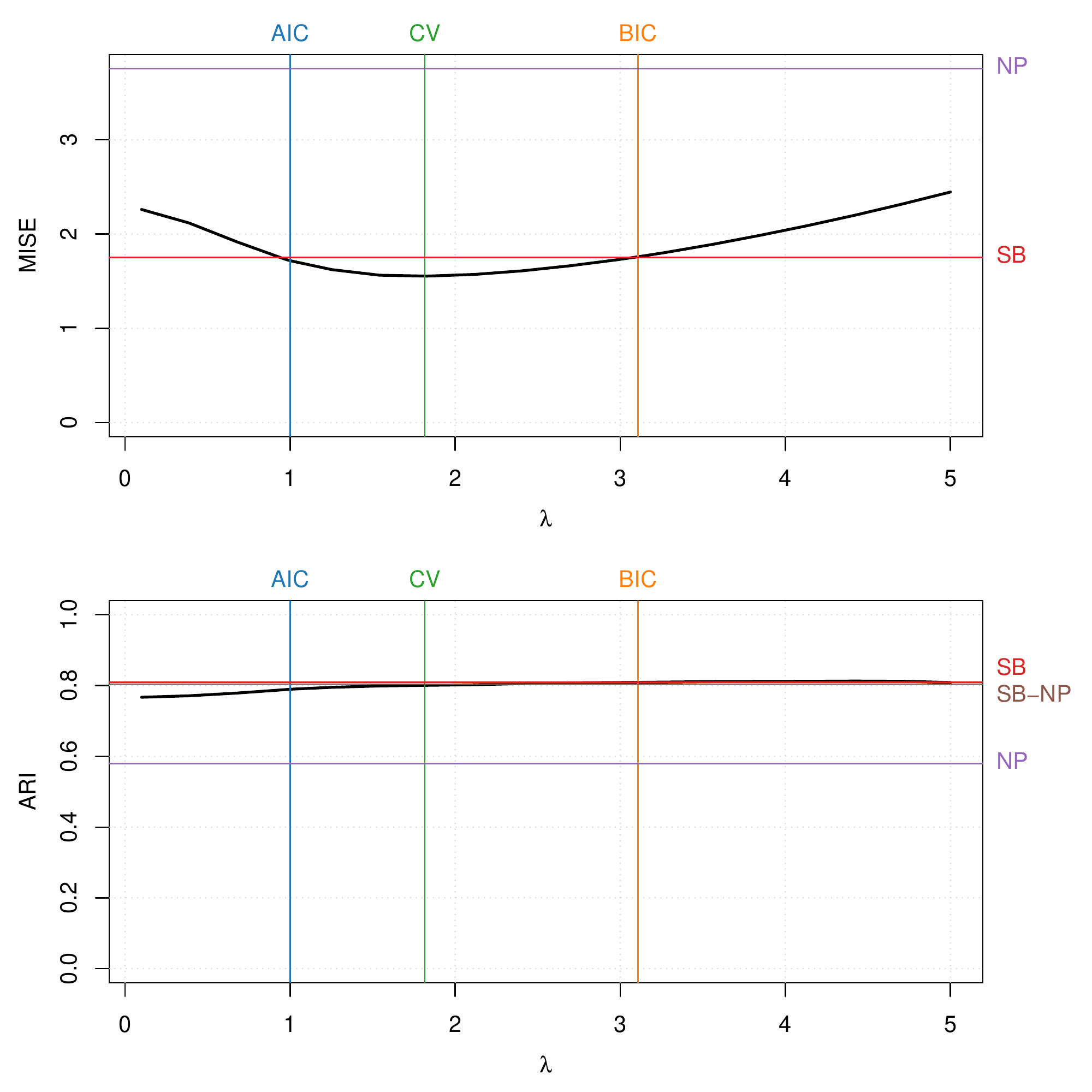}} & \multicolumn{2}{c}{\includegraphics[scale=0.23]{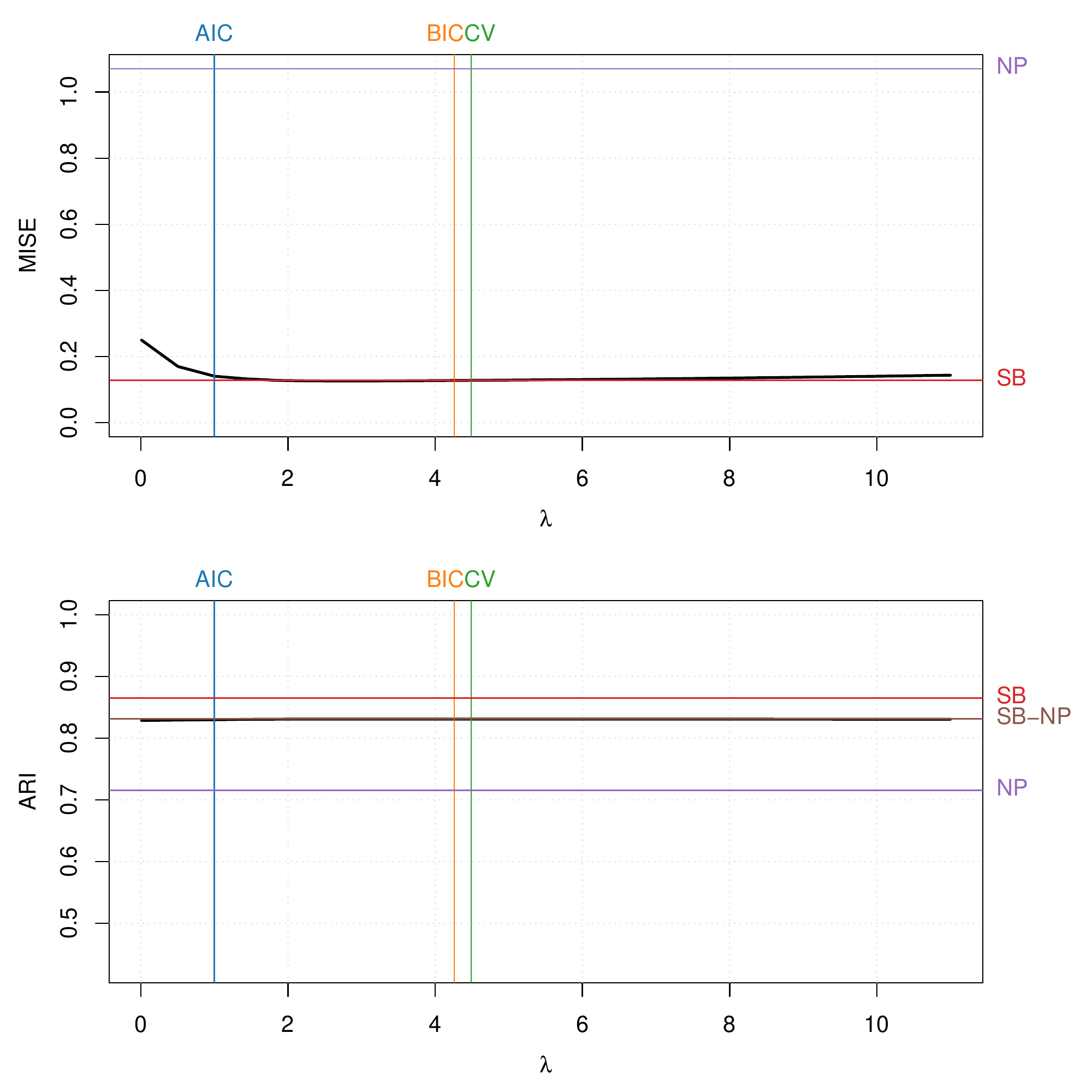}} \\ 
 \hline
  & MISE & ARI & MISE & ARI \\ \hline
 SB & 1.751 (1.387) & 0.809 (0.099) & 0.128 (0.063) & 0.865 (0.008) \\
 NP & 3.753 (0.701) & 0.580 (0.151) & 1.071 (0.191) & 0.715 (0.097) \\
 SB-NP & - & 0.804 (0.078) & - & 0.831 (0.012) \\
  $\lambda_{CV}$ & 1.586 (0.855) & 0.790 (0.075) & 0.129 (0.065) & 0.830 (0.016) \\
 $\lambda_{AIC}$ & 1.718 (0.837) & 0.790 (0.071) & 0.140 (0.064) & 0.829 (0.017) \\
 $\lambda_{BIC}$ & 1.759 (1.083) & 0.809 (0.063) & 0.128 (0.065) & 0.830 (0.012) \\
\hline

\end{tabular}
\end{center}
\caption{Cf. Table \ref{tab:M1_scrucca}. Results refer to density M3}
\label{tab:M3_scrucca}
\end{table}

\begin{table}[t!]
\begin{center}
\begin{tabular}{lcc|cc}
\hline
 & \multicolumn{2}{c|}{$n=500$} & \multicolumn{2}{c}{$n=5000$} \\
 \hline
 & \multicolumn{2}{c|}{\includegraphics[scale=0.23]{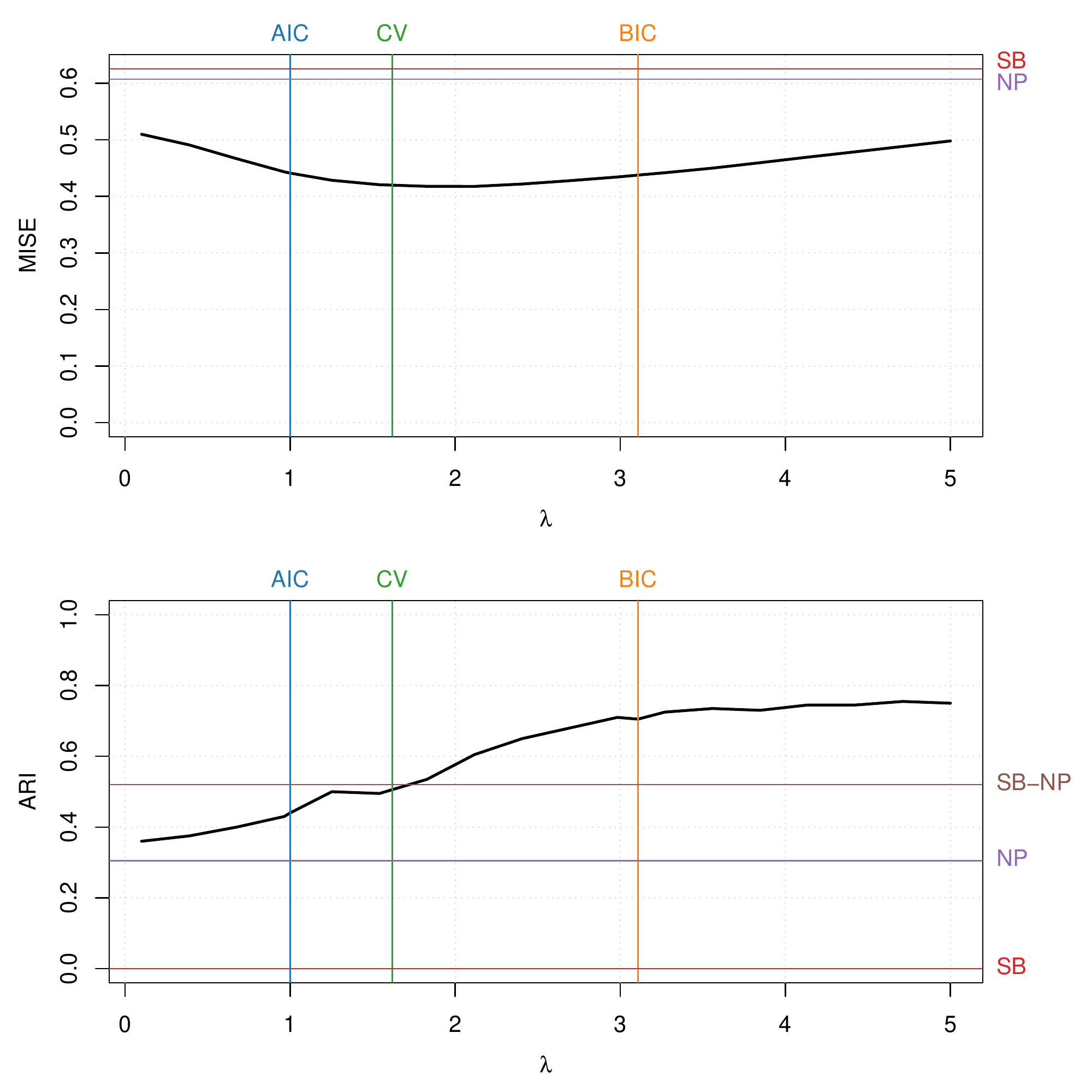}} & \multicolumn{2}{c}{\includegraphics[scale=0.23]{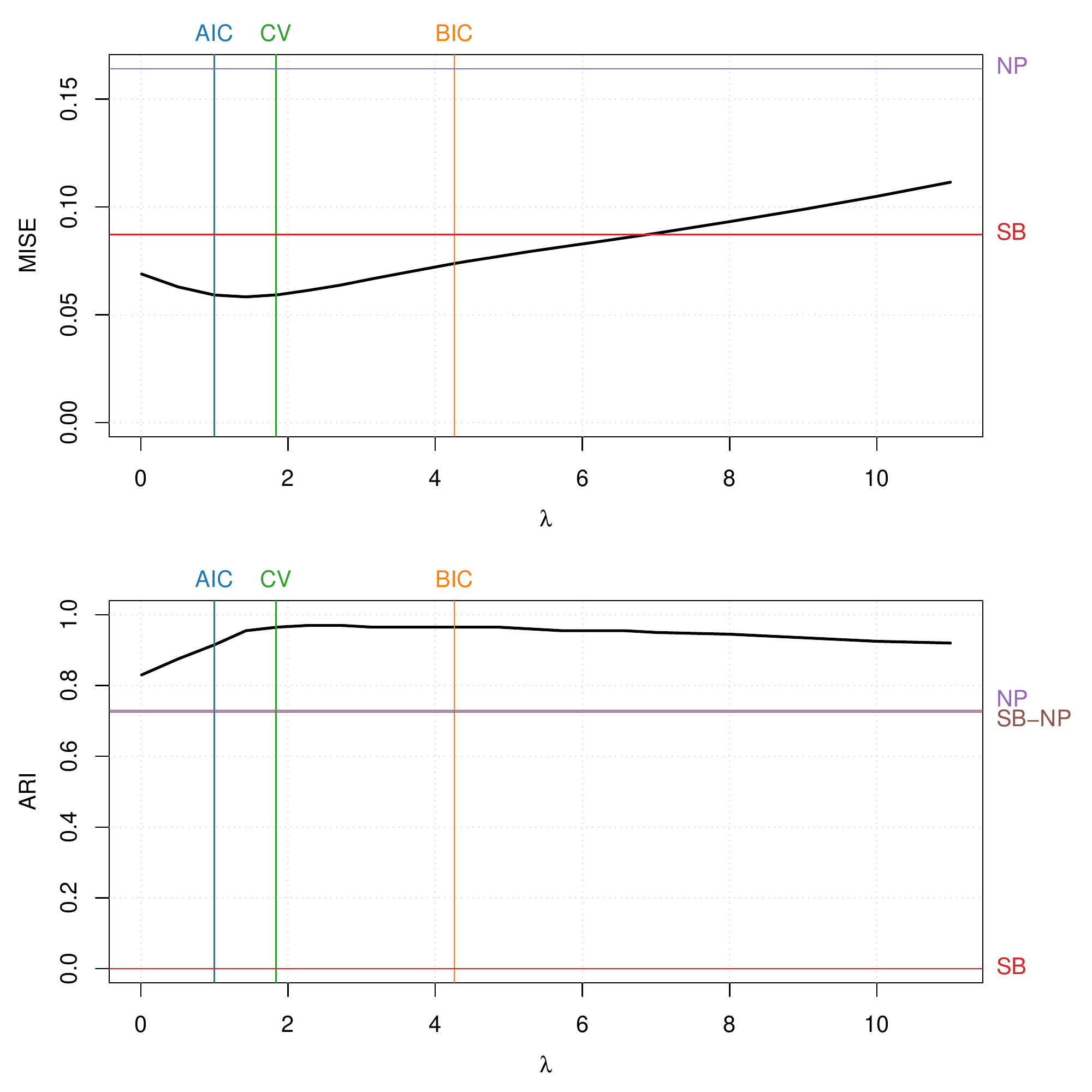}} \\ 
 \hline 
 & MISE & ARI & MISE & ARI \\ \hline
 SB & 0.626 (0.235) & 0.000 (0.000) & 0.087 (0.032) & 0.000 (0.000) \\
 NP & 0.607 (0.152) & 0.305 (0.462) & 0.164 (0.033) & 0.730 (0.445) \\
 SB-NP & - & 0.520 (0.501) & - & 0.725 (0.448) \\
  $\lambda_{CV}$ & 0.436 (0.172) & 0.440 (0.498) & 0.059 (0.026) & 0.850 (0.358) \\
 $\lambda_{AIC}$ & 0.441 (0.181) & 0.440 (0.498) & 0.059 (0.024) & 0.915 (0.280) \\
 $\lambda_{BIC}$ & 0.438 (0.167) & 0.705 (0.457) & 0.074 (0.030) & 0.965 (0.184) \\
\hline
\end{tabular}
\end{center}
\caption{Cf. Table \ref{tab:M1_scrucca}. Results refer to density M4}
\label{tab:M4_scrucca}
\end{table}

\begin{table}[t!]
\begin{center}
\begin{tabular}{lcc|cc}
\hline
 & \multicolumn{2}{c|}{$n=500$} & \multicolumn{2}{c}{$n=5000$} \\
 \hline
 & \multicolumn{2}{c|}{\includegraphics[scale=0.23]{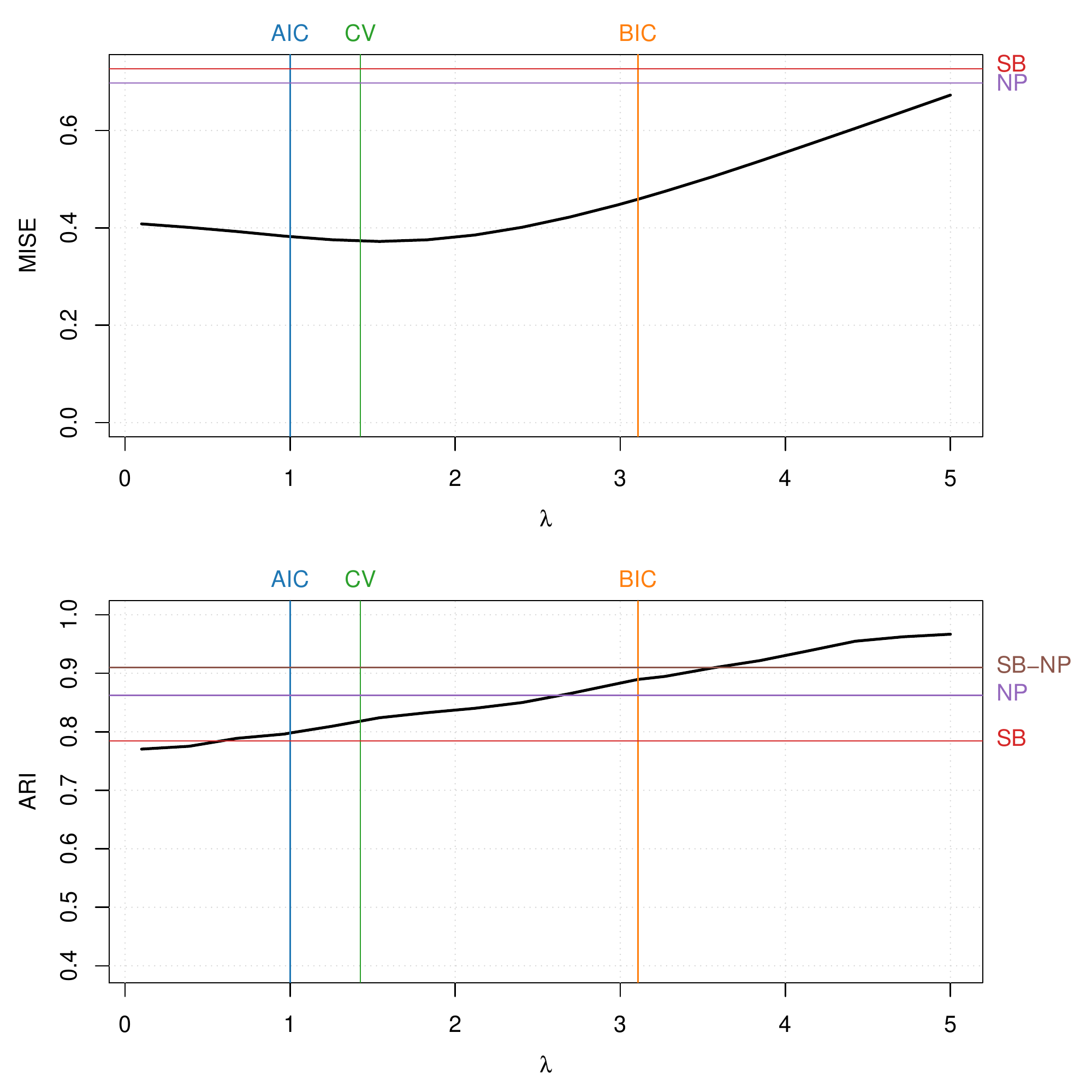}} & \multicolumn{2}{c}{\includegraphics[scale=0.23]{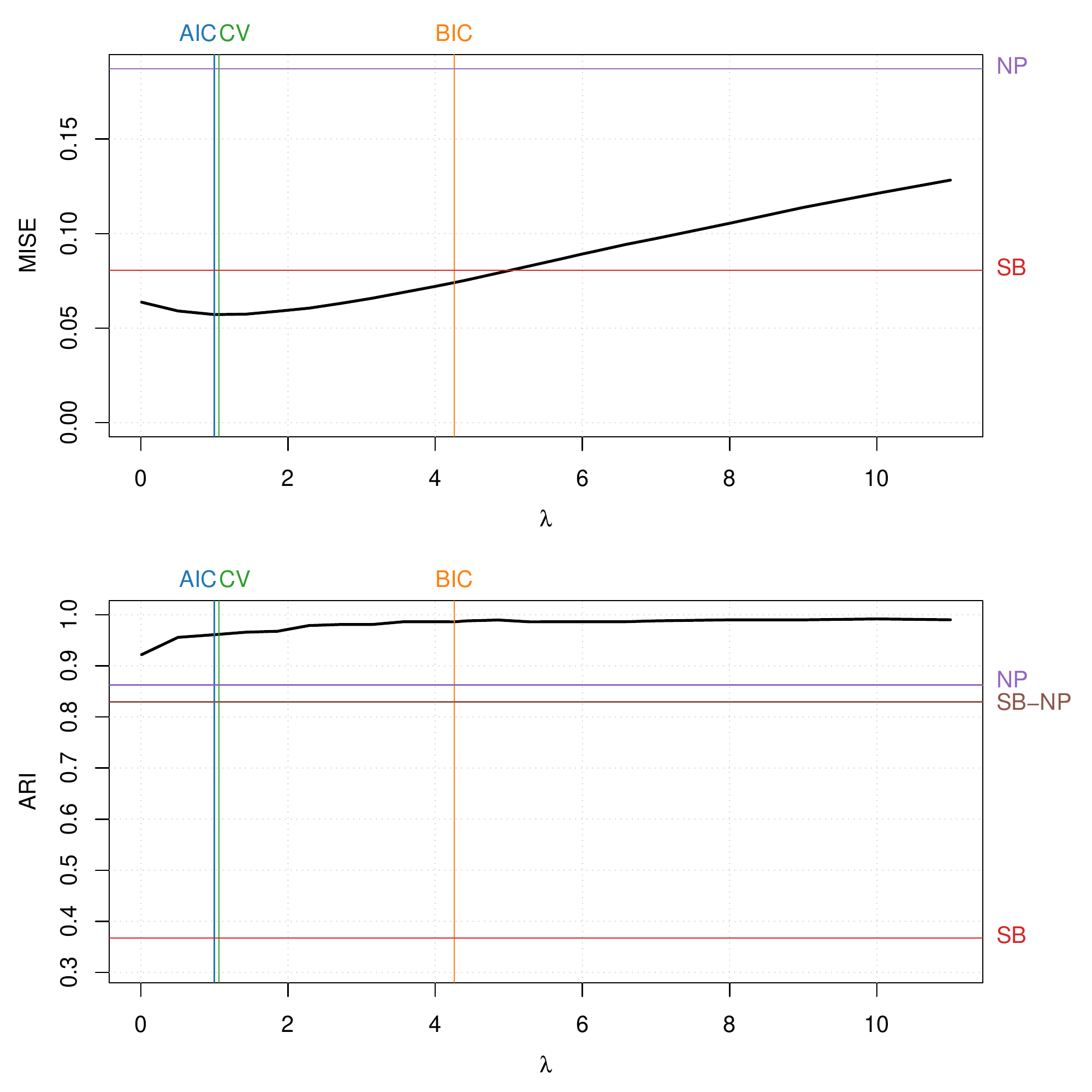}} \\ 
 \hline 
  & MISE & ARI & MISE & ARI \\ \hline
 SB & 0.727 (0.236) & 0.784 (0.193) & 0.081 (0.030) & 0.367 (0.032) \\
 NP & 0.697 (0.128) & 0.862 (0.143) & 0.187 (0.025) & 0.862 (0.164) \\
 SB-NP & - & 0.910 (0.158) & - & 0.829 (0.177) \\
  $\lambda_{CV}$ & 0.385 (0.124) & 0.789 (0.178) & 0.058 (0.014) & 0.934 (0.114) \\
 $\lambda_{AIC}$ & 0.382 (0.122) & 0.798 (0.175) & 0.057 (0.013) & 0.961 (0.091) \\
 $\lambda_{BIC}$ & 0.459 (0.139) & 0.889 (0.159) & 0.074 (0.019) & 0.986 (0.052) \\
\hline
\end{tabular}
\end{center}
\caption{Cf. Table \ref{tab:M1_scrucca}. Results refer to density M5}
\label{tab:M5_scrucca}
\end{table}

Results are reported in Tables \ref{tab:M1_scrucca} to \ref{tab:M5_scrucca}. A first expected behaviour indicates that the performances of the considered methods tend to improve with the increase of the sample size, both from a clustering and from a density estimation point of view.

Generally speaking our proposal, regardless of the penalization used, produces satisfactory density estimates and partitions of the datasets. The first three scenarios have been considered to see how the ensemble approach behaves in situations where the \emph{single best model} has a head start; in these cases the true generative model is indeed among the ones estimated in the model-based clustering routine. Even in these somewhat unfavourable settings, where in some sense an ensemble approach is not strictly needed, the proposed method behaves well producing overall comparable results with respect to the parametric ones.  

In the skewed scenarios M4 and M5, where Gaussian mixture models are known to be less effective as a clustering tool, the ensemble approach induces remarkable improvements in the performances, both in terms of MISE and ARI. Note that, regarding the relation between performances and sample size, we are witnessing some results constituting an exception with respect to what we pointed out before. Indeed, especially for the setting M5, the increased availability of data points forces Gaussian mixture models to resort to an higher number of components, even in the presence of two groups, to properly model the asymmetry thus deteriorating the clustering results. In commenting these results some words of caution are needed since obtaining the allocation according to the modal concept of groups can have a strong impact in these two settings. Nonetheless comparisons with the hybrid approach help shedding light on this and to study further the improvements intrinsically introduced by averaging together distinct densities. The method proposed, despite showing comparable results when $n=500$, attains notable enhancements when $n=5000$ along with decreased standard errors. This could constitute, from a clustering standpoint, an indication of the improved quality of the density estimates produced considering model (\ref{eq:eq1}) with respect to the ones produced by a single mixture model; better ARI values could indeed indicate smoother estimates, being easier to be explored when searching for the modes. 

The aforementioned decrease in the variability of the results of the proposal with respect to the competitors is witnessed across all the scenarios. This represents a substantial and somewhat expected advantage of the ensemble approach, since a gain in robustness and stability moves towards the desired direction when mixing models together. 

With regard to the choice of the penalization scheme some different considerations arise. As expected, building on a data-based rationale, $\lambda_{CV}$ seems to be more reliable when the aim is to obtain an accurate estimate of the density. Choosing the amount of the penalization via cross-validation appears to be particularly suitable especially when $n=500$ while, with increasing sample size, the performances of the three considered schemes tend to be more similar. However, when clustering is the final aim of the analysis $\lambda_{BIC}$ turns out to be a serious candidate as it often produces better results with respect to $\lambda_{CV}$ and $\lambda_{AIC}$; this constitutes a notable result since the \emph{BIC-type} penalization, unlike the cross-validation based one, requires a null computational cost when dealing with the selection of $\lambda$. On the other hand, not even depending on the sample size, $\lambda_{AIC}$ tends to produce the most unsatisfactory results among the three as expected. 

Lastly note that the performances of the fully nonparametric approach appear not to be competitive with the other approaches considered. Nonetheless we believe that some tuning in choosing the smoothing parameters used could lead to an improvement in the results. This is not explored in our numerical experiments since appropriate bandwidth selection is not the aim of the present study, hence it appears reasonable to resort to a standard selector as we did.


\subsection{Real data}\label{sec:chscrucca_realdata}
\noindent In this section we consider three illustrative examples on real datasets. As in the previous section, we fit our proposed model considering the three different penalization schemes introduced in Section \ref{sec:chscrucca_estimation} and we use as competitors the parametric, the nonparametric and the hybrid approaches. The number of models in the ensemble is set to $M=30$ following the same rationale as the one discussed in the simulated examples. Not having a real density to refer to, the analyses focus on the quality of the partitions obtained, evaluated via the ARI. 

\subsubsection{Iris data}\label{sec:chscrucca_iris}
\noindent The \emph{Iris} dataset (available at \texttt{https://archive.ics.uci.edu/ml/datasets/\\Iris}), already mentioned in Section \ref{sec:chscrucca_frameworkmodelspec} to motivate our proposal, have been thoroughly studied since the seminal paper by \citet{fisher1936use} and it consists in $d=4$ variables (sepal length and width, petal length and width) measured on $n=150$ iris plants with $K_{true}=3$ classes equally sized. A widely known characteristic of these data consists in having a class being linearly separable from the other two, in turn hardly to detect as separate groups. \\ Results are shown in Table \ref{tab:table_iris}. The method proposed here clearly outperforms all the considered competitors. As seen in Section \ref{sec:chscrucca_frameworkmodelspec} the BIC select a two-component model hence giving wrong indications about the number of groups. As a consequence, both the parametric and the hybrid approaches, relying on the single best model, tend to produce unsatisfactory results. On the other hand the detection of 7 groups, via modal clustering based on kernel density estimation, is a symptom of an undersmoothed density estimate with the selected bandwidth matrix. Note that the high degree of rounding in the dataset could affect nonparametric performances since the estimator is built to work with continuous data, hence without duplicated values. Our method, regardless of the penalization scheme, produces strong improvements in the clustering results. The \emph{AIC-type} and the CV-based penalties wrongly find 4 clusters with one spurious, yet small, group detected. On the contrary, a closer examination of the results reveals that $\lambda_{BIC}$ assumes roughly twice the value of $\lambda_{AIC}$ and $\lambda_{CV}$ and leads to the correct identification of 3 groups. 

\begin{table}[!t]
\centering
\begin{tabular}{rrrrrrr}
  \hline
 & SB & NP & SB-NP & $\lambda_{CV}$ & $\lambda_{AIC}$ & $\lambda_{BIC}$  \\
  \hline
 $\lambda$ & - & - & - & 1.449 & 1.000 & 2.505  \\
ARI & 0.568 & 0.556 & 0.568 &  0.869  & 0.845 & 0.941 \\ 
$\hat{K}$ & 2 & 7 & 2 & 4 & 4 & 3 \\ 
   \hline
\end{tabular}
\caption{Results obtained on the Iris dataset. The true number of cluster is $K_{true}=3$.}
\label{tab:table_iris}
\end{table}

\subsubsection{DLBCL data}\label{sec:chscrucca_dlbcl}
\begin{figure}[!t]
\center
\includegraphics[scale=0.24]{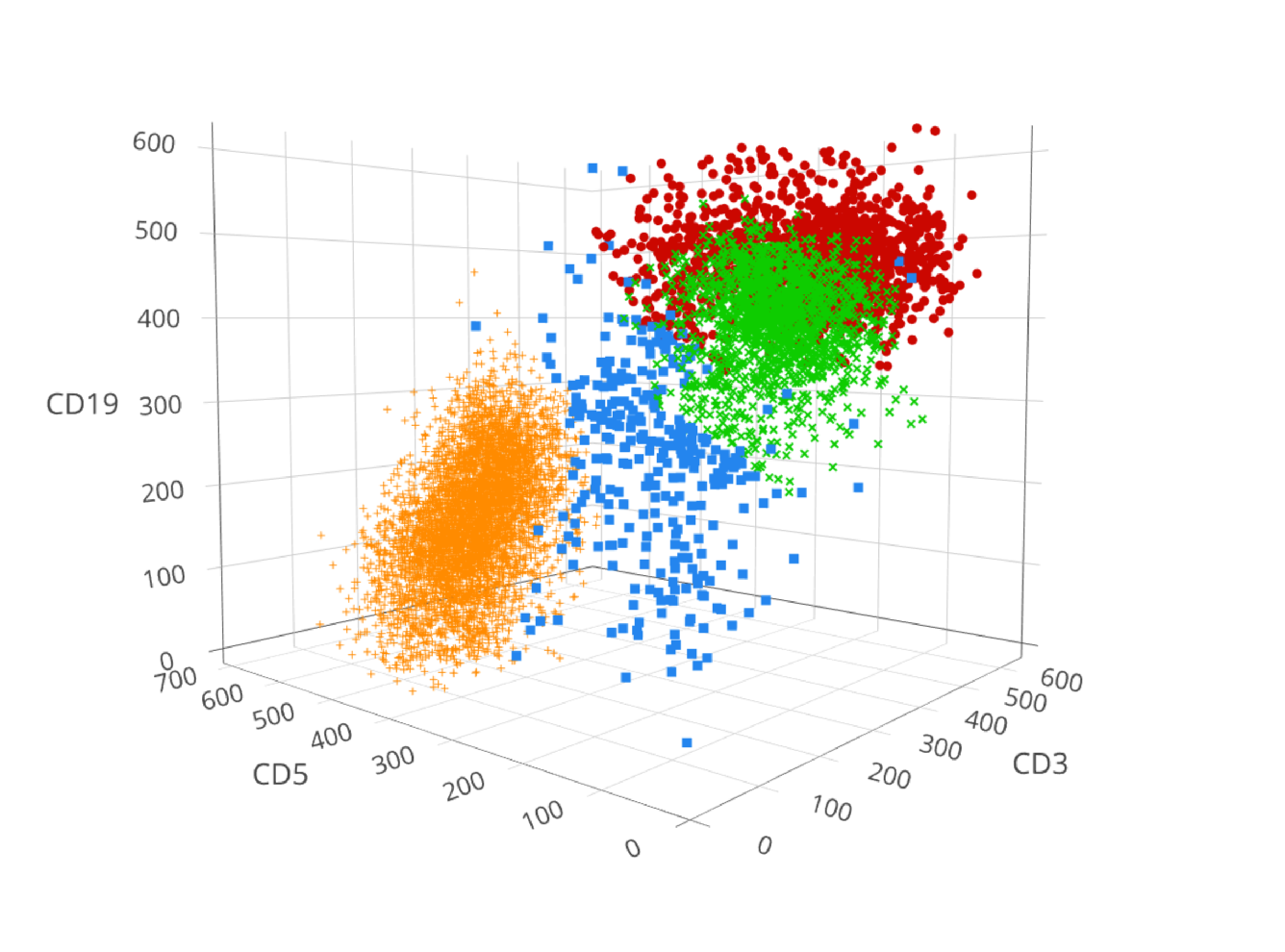}
\caption{3D scatter plot of the DLBCL data with colors representing the true clustering labels.}
\label{fig:contourdlbcl}
\end{figure}

\noindent The \emph{Diffuse Large B-cell Lymphoma} (DLBCL) dataset is provided by the British Columbia Cancer Agency \citep{spidlen2012flowrepository,aghaeepour2013critical}. The sample consists in fluorescent intensities of $d=3$ markers, namely CD3, CD5 and CD19, measured on $n=8183$ lymph nodes cells from subjects with a DLBCL diagnosis. A scatter plot of the data is shown in Figure \ref{fig:contourdlbcl}. In flow cytometry analysis these measurements are used to study normal and abnormal cell structures and to monitor human diseases and response to therapies. An essential step in this framework consists in obtaining a grouping of the cells according to their fluorescences. This task is usually accomplished via the so called \emph{gating} process: the experts obtain a partition manually by visually inspecting the data. This approach is usually time-consuming and infeasible in high-dimensional situations, therefore clustering tools could come in aid to automate the gating process. 
The 3-dimensional structure of the data, illustrated in Figure \ref{fig:contourdlbcl}, allows us to visually inspect the true cluster configuration, displaying elongated and skewed group shapes. As noted in the simulated scenarios, results in Table \ref{tab:table_dlbcl} show how the model-based approach by using symmetric components tends to perform badly when dealing with such situations, since it detects an higher number of groups with respect to the true one. In this setting, building mixtures on more flexible, possibly skew component densities could help in improving the fit by means of a single model. Conversely, the nonparametric and the hybrid approaches, which search for the modes of the density, do not suffer of the same drawbacks and outperform the parametric strategy. Nonetheless, while the former appears to undersmooth again the density, the latter detects the true number of clusters, yet with improved performance in the allocation of units.

Our proposal, regardless of the penalization scheme adopted, enjoys the very same advantage of modal clustering methods when dealing with asymmetric shapes. In fact the results obtained improve with respect to the hybrid approach thus indicating that our model produces a density estimate better tailored for the clustering scope. In this case different penalization schemes lead to irrelevant changes in the ARI values, and indicate a weaker dependency on the selected penalty value. 

\begin{table}[t]
\centering
\begin{tabular}{rrrrrrr}
  \hline
 & SB & NP & SB-NP & $\lambda_{CV}$ & $\lambda_{AIC}$ & $\lambda_{BIC}$ \\ 
  \hline
$\lambda$ & - & - & - & 0.001 & 1.000 & 4.505 \\
ARI & 0.401 & 0.857 & 0.867 & 0.912 & 0.909  & 0.910 \\ 
$\hat{K}$ & 7 & 5 & 4 & 4 & 4 & 4 \\ 
   \hline
\end{tabular}
\caption{Results obtained on the DLBCL dataset. The true number of cluster is $K_{true}=4$.}
\label{tab:table_dlbcl}
\end{table}

\subsubsection{Olive oil data}\label{sec:chscrucca_olive}
\noindent As a last example we consider the \emph{Olive oil} dataset, originally introduced in \citet{forina1986multivariate}. The data consist of $d=8$ chemical measurements on $n=572$ olive oils produced in 9 regions of Italy (North and South Apulia, Calabria, Sicily, Sardinia coast and inland, Umbria, East and West Liguria) that can be further aggregated in three macro-areas (Centre-North, South and Sardinia island). Clustering tools may come in aid in reconstructing the geographical origin of the oils on the basis of their chemical compositions. 

\begin{table}[t]
\centering
\begin{tabular}{rrrrrrr}
  \hline
 & SB & NP & SB-NP & $\lambda_{CV}$ & $\lambda_{AIC}$ & $\lambda_{BIC}$ \\ 
  \hline
  $\lambda$ & - & - & - &1.155 & 1.000 & 3.175\\
ARI & 0.782 & 0.604 & 0.792 & 0.902 & 0.902 & 0.892 \\ 
$\hat{K}$ & 6 & 20 & 6 & 8 & 8 & 8 \\ 
   \hline
\end{tabular}
\caption{Results obtained on the Olive oil dataset. The unaggregated regions have been considered as true labels hence $K_{true}=9$.}
\label{tab:table_olive}
\end{table}

Compared to the cases considered previously, this example allows exploring the performances of the proposal in a moderately higher dimensional setting. Results in Table \ref{tab:table_olive} show how our proposal outperforms the competitors, regardless of the penalization adopted, and how it yields a more faithful partition of the data into the 9 considered regions. The parametric and the hybrid approaches detect 6 groups, aggregating Sardinia coast and inland oils and highlighting some issues concerning the correct classification of oils produced in South macro-area. On the other hand, probably suffering of the higher dimensionality of the data, the fully nonparametric approach clearly produces a partition based on an severely undersmoothed density with 20 modes. 

As it happened in Section \ref{sec:chscrucca_dlbcl} the clustering performances of our proposal appear to be quite insensitive to the specific penalization adopted. In Table \ref{tab:olive_res_lambda_aic} we report the partition induced by considering $\lambda_{AIC}$ as penalizing parameter. Again it appears harder to discriminate the oils produced in the southern macro-area, with calabrian and sicilian ones assigned mainly to the same cluster, while oils in the other two macro-areas are substantially correctly identified. \\

\begin{table}[t]
\begin{center}
\begin{tabular}{llcccccccc}
& & 1 & 2 & 3 & 4 & 5 & 6 & 7 & 8 \\
    \hline
    \multirow{4}{*}{South}& Apulia north & 24 & 1 & 0 & 0 & 0 & 0 & 0 & 0 \\
    & Apulia south & 0 & 6 & 200 & 0 & 0 & 0 & 0 & 0  \\
    & Calabria & 0 & 56 & 0 & 0 & 0 & 0 & 0 & 0 \\
    & Sicily & 6 & 30 & 0 & 0 & 0 & 0 & 0 & 0\\
    \hline
    \multirow{2}{*}{Sardinia} & Sardinia inland & 0 & 0 & 0 & 65 & 0 & 0 & 0 & 0 \\
    & Sardinia coast & 0 & 0 & 0 & 0 & 33 & 0 & 0 & 0 \\
    \hline
    \multirow{3}{*}{Centre-North} & Liguria east & 0 & 0 & 0 & 0 & 0 & 1 & 42 & 7 \\
    & Liguria west & 0 & 0 & 0 & 0 & 0 & 0 & 0 & 50 \\
    & Umbria & 0 & 0 & 0 & 0 & 0 & 48 & 3 & 0 \\
    \hline
\end{tabular}
\end{center}
\caption{Olive oil results, partition obtained with penalization parameter $\lambda_{AIC}$}
\label{tab:olive_res_lambda_aic}
\end{table}

\section{Conclusions}\label{sec:chscrucca_conclusions}

In this work we have addressed the issue of  overcoming the strong reliance of model-based clustering on a single best model, selected according to some information criterion. Making reference to a single model may be suboptimal both for clustering and for density estimation, since alternative well-fitted models may provide useful information by uncovering different and complementary features which are otherwise discarded. It has been pointed out that possible solutions may be found in the ensemble learning literature. In this setting, we have proposed a clustering method 
building on a density function which averages different estimated models, and whose modal regions are then associated to the groups. 
The introduced density estimator is defined as a convex linear combination of the estimates of the models in the ensemble, with weights estimated via penalized maximum likelihood. This choice allows assigning relevance to the only models which better fit the data while avoiding the risk of overfitting. 

The proposed methodology finds a relevant strength in the coherency not to resort to distance-based approaches to practically identify a grouping of the data. While it exhibits an hybrid taxonomy which borrows both concepts and operational tools from the parametric and nonparametric settings, it enjoys some of the relevant properties of any formulation of density-based clustering, as for example its mathematically sound formalization. In fact, it additionally inherits the strengths of both approaches. On one hand, by resorting to parametric tools and to model average, density estimation is strengthened and allows obtaining more accurate results of both nonparametric tools and single parametric models. Additionally, the introduced penalization scheme allows us to end up with a single parsimonious mixture model when the true generative mechanism underlying the data is among the candidates to be averaged. On the other hand, relying on the modal concept of clusters, groups are not constrained by any shape induced by distributional assumptions, and they can arise with arbitrary shapes which naturally comply with the geometric intuition. 

The introduced method finds its original and main motivation in the need of proposing a viable alternative to the single best model paradigm in the model-based clustering framework. Nonetheless a different reading key may be given if the proposal is considered from a modal perspective. Nonparametric clustering is known to have its Achilles' heel in the density estimation step, since nonparametric estimators tend not to be reliable in moderately high-dimensional situations \citep{scott}. It has been shown \citep{fraley2002model} that Gaussian mixture models, used with density estimation purposes, usually outperform kernel density estimator even in low-dimensional spaces. Our proposed method may therefore be thought as a way to strengthen the estimation process, by using an ensemble of Gaussian mixtures, with a modal clustering aims in mind.

It is worth noting that the proposed density estimator lies itself half-away between parametric and nonparametric methods. In fact, when considering the number of components as an unknown parameter, mixture models can be seen as a semi-parametric compromise between classical parametric models and nonparametric methods as kernel density estimators. The model we introduced has an increased number of components with respect to a single mixture, inherited by the averaging procedure, hence it takes another step forwards the nonparametric approach to density estimation. 

The performances of the proposal have been investigated both on simulated and on real data, selected to encompass different situations and confirm the considerations above. The method produces satisfactory results both from a density estimation and from a clustering perspective, and it compares favorably with the considered competitors. A deeper examination of the results leads to disentangle the reasons of the improvements into two different sources: on one side partitioning the data according to the modal formulation produces promising results in some specific scenarios, on the other hand several clues have been obtained which highlight enhancements in the density estimation process. Moreover, since nonparametric density estimation performances are known to deteriorate in high-dimensional settings, our proposal is expected to produce more pronounced improvements in these scenarios. Even if in the simulation study we explored only two-dimensional situations due to the computational burden, some analysis we have conducted, not reported here, appears to confirm this conjecture. 

Concerning the introduced penalization schemes, the results seem to suggest the use of the \emph{BIC-type} penalization, being more suitable for clustering, or of the cross-validation-based one, being able to adapt more to the features of the considered dataset. Lastly note that the use of a penalized log-likelihood, as well as the ways we considered to select the penalization parameter, targets an improvement in density estimation. While the aforementioned considerations motivate the soundness of this choice, a possible interesting research direction may consist in proposing some clustering-oriented penalization schemes and comparing them with the one introduced in this work.


\appendix 
\section{Parameter settings}\label{App:settings}

In the following the parameter settings of the densities selected for the simulations in Section \ref{sec:chscrucca_simulation} are presented. For Density M1, M2 and M3, being Gaussian mixture models, we adopt the usual notation where, for a given $k$ component, $\pi_k$ represents the $k$th mixture weight, $\mu_k$ and $\Sigma_k$ the corresponding mean vector and covariance matrix. On the other hand, for Density M4 and M5 we consider multivariate skew normal distributions (or mixture of) hence the additional parameter $\delta_k$ regulates the skeweness of the $k$th component \citep[for details on the parametrization readers can refer to][]{azzalini1996multivariate}.

\subsection*{Density M1}
\vspace*{-0.8cm}
\begin{table}[H]
\center
\begin{tabular}{ccccc}
\toprule
 & Component & $\pi_k$ & $\mu_k$ & $\Sigma_k$  \\
 \midrule
  & 1 & 1 & $\begin{pmatrix} 0 \\ 0 \end{pmatrix}$ & $\begin{pmatrix} 1.25 & 0.75 \\ 0.75 & 1.25  \end{pmatrix}$ \\
 \bottomrule
\end{tabular}
\end{table}

\subsection*{Density M2}
\vspace*{-0.8cm}
\begin{table}[H]
\center
\begin{tabular}{ccccc}
\toprule
 & Component & $\pi_k$ & $\mu_k$ & $\Sigma_k$  \\
  \midrule
  & 1 & 0.5 & $\begin{pmatrix} -0.53 \\ -0.53 \end{pmatrix}$ & $\begin{pmatrix} 0.68 & -0.41 \\ -0.41 & 0.68 \end{pmatrix}$ \vspace*{0.2cm} \\ 
 & 2 & 0.5 & $\begin{pmatrix} 0.53 \\ 0.53 \end{pmatrix}$ & $\begin{pmatrix} 0.68 & -0.41 \\ -0.41 & 0.68 \end{pmatrix}$ \\
 \bottomrule
\end{tabular}
\end{table}

\subsection*{Density M3}
\vspace*{-0.8cm}
\begin{table}[H]
\center
\begin{tabular}{ccccc}
\toprule
 & Component & $\pi_k$ & $\mu_k$ & $\Sigma_k$  \\
  \midrule
  & 1 & 0.4 & $\begin{pmatrix} -0.85 \\ -0.85 \end{pmatrix}$ & $\begin{pmatrix} 0.58 & -0.35 \\ -0.35 & 0.58  \end{pmatrix}$ \vspace*{0.2cm} \\
 & 2 & 0.4 & $\begin{pmatrix} 0.85 \\ 0.85 \end{pmatrix}$ & $\begin{pmatrix} 0.58 & -0.35 \\ -0.35 & 0.58  \end{pmatrix}$ \vspace*{0.2cm} \\
 & 3 & 0.2 & $\begin{pmatrix} 0 \\ 0 \end{pmatrix}$ & $\begin{pmatrix} 0.16 & -0.09 \\ -0.09 & 0.16  \end{pmatrix}$ \\
 \bottomrule
\end{tabular}
\end{table}

\subsection*{Density M4}
\vspace*{-0.8cm}
\begin{table}[H]
\center
\begin{tabular}{cccccc}
\toprule
 & Component & $\pi_k$ & $\mu_k$ & $\Sigma_k$ & $\delta_k$ \\
 \midrule
  & 1 & 1 & $\begin{pmatrix} 0 \\ 0 \end{pmatrix}$ & $\begin{pmatrix} 0.8 & -0.4 \\ -0.4 & 0.8  \end{pmatrix}$ & $\begin{pmatrix} 3 \\ 3 \end{pmatrix}$  \\
 \bottomrule
\end{tabular}
\end{table}

\subsection*{Density M5}
\vspace*{-0.8cm}
\begin{table}[H]
\center
\begin{tabular}{cccccc}
\toprule
 & Component & $\pi_k$ & $\mu_k$ & $\Sigma_k$ & $\delta_k$ \\
 \midrule
  & 1 & 0.5 & $\begin{pmatrix} 1 \\ 1 \end{pmatrix}$ & $\begin{pmatrix} 0.8 & -0.4 \\ -0.4 & 0.8  \end{pmatrix}$ & $\begin{pmatrix} 3 \\ 3 \end{pmatrix}$  \vspace*{0.2cm} \\
  & 2 & 0.5 & $\begin{pmatrix} -1 \\ -1 \end{pmatrix}$ & $\begin{pmatrix} 0.8 & -0.4 \\ -0.4 & 0.8  \end{pmatrix}$ & $\begin{pmatrix} -3 \\ -3 \end{pmatrix}$  \\
\bottomrule
\end{tabular}
\end{table}

\section*{Conflict of interest}
The authors declare that they have no conflict of interest.

\bibliographystyle{spbasic}      
\bibliography{biblio}   

\end{document}